\documentclass[amsfonts,12pt]{article}

\usepackage{amsmath, amssymb, amsfonts, amsthm}
\title{}\date{}

\newcount\hour \newcount\minute
\hour=\time  \divide \hour by 60
\minute=\time
\loop \ifnum \minute > 59 \advance \minute by -60 \repeat
\def\nowtwelve{\ifnum \hour<13 \number\hour:
                      \ifnum \minute<10 0\fi
                      \number\minute
                      \ifnum \hour<12 \ A.M.\else \ P.M.\fi
         \else \advance \hour by -12 \number\hour:
                      \ifnum \minute<10 0\fi
                      \number\minute \ P.M.\fi}
\def\nowtwentyfour{\ifnum \hour<10 0\fi
                \number\hour:
                \ifnum \minute<10 0\fi
                \number\minute}

\title{Explicit Integration of Friedmann's Equation \\ with   Nonlinear
Equations of State}

\author{Shouxin Chen\footnote{Email address: chensx@henu.edu.cn}\\Institute of Contemporary Mathematics\\School of Mathematics\\Henan University\\
Kaifeng, Henan 475004, PR China\\\\
Gary W. Gibbons\footnote{Email address: gwg1@damtp.cam.ac.uk}\\
D. A. M. T. P.\\
University of Cambridge\\
Cambridge CB3 0WA, U. K.\\\\Yisong Yang\footnote{Email address: yisongyang@nyu.edu}\\Department of Mathematics\\Polytechnic School, New York University\\Brooklyn, New York 11201, U. S. A\\ \&\\NYU-ECNU 
Institute of Mathematical Sciences\\New York University - Shanghai\\3663 North Zhongshan Road, Shanghai 200062, PR China}
\newcommand{\bfR}{{\Bbb R}}
\newcommand{\arctanh}{{\mbox{arctanh}}}

\def\ben{\begin{equation}}
\def \een{\end{equation}} 
\def \half {\frac{1}{2}}

\def\XXint#1#2#3{{\setbox0=\hbox{$#1{#2#3}{\int}$}
 \vcenter{\hbox{$#2#3$}}\kern-.5\wd0}}

\newtheorem{oldtheorem}{Theorem}[section]
\newtheorem{oldassertion}[oldtheorem]{Assertion}
\newtheorem{oldproposition}[oldtheorem]{Proposition}
\newtheorem{oldremark}[oldtheorem]{Remark}
\newtheorem{oldlemma}[oldtheorem]{Lemma}
\newtheorem{olddefinition}[oldtheorem]{Definition}
\newtheorem{oldclaim}[oldtheorem]{Claim}
\newtheorem{oldcorollary}[oldtheorem]{Corollary}

\newcommand{\dd}{\mbox{d}}
\newcommand{\ee}{\end{equation}}
\newcommand{\be}{\begin{equation}}
\newcommand{\bea}{\begin{eqnarray}}
\newcommand{\eea}{\end{eqnarray}}

\newcommand{\e}{\mbox{e}}
\newcommand{\pa}{\partial}
\newcommand{\Om}{\Omega}

\newcommand{\nn}{\nonumber}

\newcommand{\lm}{\lambda}
\def \cE{{\cal E}} 
\begin{document}
\maketitle


\begin{abstract}

In this paper we study the integrability of the Friedmann equations, when the equation of state for the perfect-fluid universe is nonlinear, in the light of  the Chebyshev theorem.
A series of important, yet not previously touched, problems will be worked out which include
the generalized Chaplygin gas, two-term energy density, trinomial Friedmann, Born--Infeld,  two-fluid models, and Chern--Simons modified gravity theory models.  With the explicit integration, we are able to
understand exactly the roles of the  physical parameters in various models play in the cosmological evolution
which may also offer clues to a profound understanding of the problems in general settings. For example, in the Chaplygin gas universe, a few integrable cases lead us to derive
a universal formula for the asymptotic exponential growth rate of the scale factor, of an explicit form, whether the Friedmann equation is integrable or not, which reveals the coupled roles played by various physical sectors and
it is seen that, as far as there is a tiny presence of  nonlinear matter, conventional linear matter makes contribution to the dark matter, which becomes significant near
the phantom divide line.
The Friedmann  equations also arise in
areas of physics not directly related to cosmology. We provide some 
examples ranging from  geometric optics and  central orbits to  soap films and
the shape of glaciated valleys to  which our results may be applied.

\medskip


\medskip


\end{abstract}

\maketitle

\tableofcontents

\section{Introduction}

The Einstein equations of general relativity reduce,
 when the spacetime metric is of the
Friedmann--Lema\^{i}tre--Robertson--Walker type governing an isotropic and homogeneous universe filled with a perfect fluid
characterized by its pressure $P_m$ and energy density $\rho_m$, to
the  Friedmann equations: a set of nonlinear 
ordinary differential equations, determining the law of evolution of the
spatial scale factor $a(t)$ , in terms of the  Hubble ``constant'',
$H= \frac{\dot a}{a}$.  It is a  challenging task, not always possible, to
obtain a complete explicit integration in finite terms.
In the recent work \cite{CGLY}, we demonstrated a link between  
a theorem of  Chebyshev   and the explicit integration
in both cosmological time $t$ and conformal time $\eta$ of the Friedmann equations
in all dimensions and with an arbitrary cosmological constant $\Lambda$.
We were able to establish that for spatially  flat universes  an explicit
integration
in $t$ may always be carried out, and that,
in the non-flat situation and when $\Lambda$ is zero
and
the ratio $w$ of $P_m$ and $\rho_m$ in the barotropic equation of state of the perfect-fluid universe is rational,  an explicit integration may be carried out if and only if the dimension $n$ of space and $w$ obey some specific relations among an infinite family. Besides, we proved that the situation for an explicit integration in $\eta$ is complementary to
that in $t$ so that in the flat-universe case
with $\Lambda\neq0$ an explicit integration in $\eta$  can
be carried out if and only if $w$ and $n$ obey similar
relations among a well-defined  family,
and that, when $\Lambda=0$, an explicit
 integration can always be carried out whether the
space is flat, closed, or open. Furthermore,
we illustrated with an example that our method is also useful in the study of more realistic
cosmological situations when the equation of state is nonlinear. The purpose of our current paper to explore the method further to get exact integrations of
some other important cases which are not covered in \cite{CGLY} and give a few examples whose integrability is beyond the reach of
the Chebyshev theorem but can be obtained by other means. We shall also see how the integration helps understand some general situations when integration is not
possible.

When the equation of state of the perfect-fluid universe is
nonlinear, the situations include the generalized Chaplygin gas, 
two-term energy density, the trinomial Friedmann, \footnote{
That is, the energy density considered as a function
of the scale factor contains three  terms, each a power of the scale factor.}
Born--Infeld, and two-fluid models. Specifically, for the
generalized Chaplygin gas model, we work on the flat-universe situation with zero cosmological constant and identify all integrable cases. For the two-term
energy density model, we can do the same. In both situations the Chebyshev theorem may be applied directly. In the trinomial Friedmann equation model, we show
that the Chebyshev theorem is applicable only in the bottom three-dimensional space case. For the Born--Infeld type fluid model and assuming a flat universe, we show that the Chebyshev theorem works only in the critical coupling situation when the cosmological constant and Newton constant fulfill a specified condition. However, in all non-critical cases, we show that the Friedmann equation allows an integration for any value of the cosmological constant.
Moreover, we study a two-fluid model and carry out its integration by the Chebyshev theorem for a closed universe situation. Its interest is that in the original
setting the integrand is not of a binomial type but it may be recast and decomposed into a binomial form so that the Chebyshev theorem is applicable. 
We will also conduct a study of the two-fluid model in terms of the reduced temperature, given as the inverse of the scale factor. In this situation the Friedmann
equation in conformal time does not allow an application of the Chebyshev theorem because it cannot be reduced into a binomial form but may still be integrated
explicitly when the cosmological constant vanishes and universe is either closed or open (the flat case is trivial). The necessity of introducing several layers of
intermediate variables in the process of integration often makes it complicated and cumbersome to express the 
 final dependence of the scale factor on either
cosmic or conformal time transparently, although such a task is always manageable. As  illustrations, we will work out 
concrete examples for the models which are integrated in terms of cosmic time. These include the Chaplygin gas, Born--Infeld, two-term energy, two-fluid, 
trinomial Friedmann equation, and Chern--Simons modified gravity theory models. Although these examples are of different technical features, the common lesson gained is that  integration allows us to
obtain both qualitatively and quantitatively accurate knowledge about the solutions, especially regarding the roles played by various physical parameters.
It should be emphasized that the study on integrable cases may often shed profound insight into  nonintegrable cases as well. Here we carry out such a study for
the important Chaplygin gas model. We will see that some integrable examples lead us to uncover a universal asymptotic exponential growth rate formula for the
expansion of the scale factor regardless whether the equation is integrable. The importance of this formula is that it reveals explicitly the roles played by various
coupling parameters
and indicates that, no matter how
small the nonlinear Chaplygin component is, the conventional linear component always join force to contribute to the exponential rate, or dark matter, even though the cosmological
constant is absent.
Moreover, this universal formula indicates that, near the so-called phantom divide line, the linear component contributes to dark matter significantly.
Finally, we will also present a collection of examples of the
Friedmann type equations which may be integrated in view of the Chebyshev theorem but fall out of the subject of relativistic cosmology.

The plan of the paper is as follows.
In Section 2  we introduce the Friedmann equation
and the relevance of the Chebyshev theorem to its integration.
The equation is usually expressed in terms of the scale
factor $a(t)$, but as, we point out,  is sometimes advantageous 
to express the scale factor as a function of cosmic time
or to express the so-called reduced temperature or inverse of the scale factor 
in terms of conformal time. In Section 3 we review some material
about perfect fluids in cosmology. In particular we show how
starting from an equation of state, we may always express the energy density 
as a function of the scale factor and conversely 
to obtain the equation  of state, given the energy density as a function
of the scale factor. We also review 
the Lagrangian approach to perfect fluids and its relation
to $k$-essence and  similar models of current interest.  
In Section 4 we  consider in detail  the 
Chaplygin gas and the Born--Infeld models.
Section 5 is concerned with various fluids with multiple
components  of the type described above. In Section
6 we discuss the integration of the reduced temperature form of the Friedmann
equation. Finally, in Section 7,  we point out various
occurrences of the Friedmann equation in  non-cosmological contexts
and  the relevance of our results to them. In Section 8 
we provide a conclusion and indicate possible future directions.


\section{Friedmann's Equation and Chebyshev's Theorem}
\label{Friedmann}
\setcounter{equation}{0}

As in \cite{CGLY}, the universe we consider here is an $(n+1)$-dimensional  homogeneous and isotropic
Lorentzian
spacetime with the metric
\be \label{1}
\dd s^2=g_{\mu\nu}\dd x^\mu\dd x^\nu=-\dd t^2+a^2(t)g_{ij} \dd x^i\dd x^j,\quad i,j=1,\dots,n,
\ee
where $t$ is the cosmological (or cosmic) time and $g_{ij}$ is the  metric of an $n$-dimensional Riemannian manifold
$M$ of constant scalar curvature characterized by an indicator, $k=-1,0,1$,
so that $M$ is an $n$-hyperboloid, the flat space $\bfR^n$, or an $n$-sphere, with the
respective metric
\be \label{2}
g_{ij}\dd x^i\dd x^j=\frac1{1-kr^2}\,\dd r^2+r^2\,\dd\Om^2_{n-1},
\ee
where $r>0$ is the radial variable and $\dd\Om_{n-1}^2$ denotes the canonical metric of the unit
sphere $S^{n-1}$ in $\bfR^n$. Inserting the metric (\ref{1})--(\ref{2}) into the Einstein equations
\be
G_{\mu\nu}+\Lambda g_{\mu\nu}=8\pi G_n T_{\mu\nu},
\ee
where $G_{\mu\nu}$ is the Einstein tensor, $G_n$ the universal gravitational constant in $n$ spatial dimensions, and $\Lambda$ the cosmological constant, the speed of light is set to unity, and $T_{\mu\nu}$ is the
energy-momentum tensor of an ideal cosmological fluid given by
\be \label{4}
T_{\mu}^{\nu}=\mbox{diag}\{-\rho_m,P_m,\dots,P_m\},
\ee
with $\rho_m$ and $P_m$ the $t$-dependent matter energy density and pressure,
we arrive after some standard reduction 
at the Friedmann equation
 \be
H^2=\frac{16\pi G_n}{n(n-1)}\rho-\frac k{a^2},\label{a1.1}
\ee
where
$H=\frac{\dot{a}}a
$
is the usual Hubble parameter, $\dot{f}=\frac{\small{\dd} f}{\small{\dd} t}$,  and $\rho,P$ are the effective energy
density and pressure given by
\be \label{8}
\rho=\rho_m+\frac{\Lambda}{8\pi G_n},\quad P=P_m-\frac{\Lambda}{8\pi G_n},
\ee
subject to the energy-conservation law
\be\label{1.2}
\dot{\rho}+n(\rho+P)H=\dot{\rho}_m+n(\rho_m+P_m)H=0.
\ee

\subsection{Reduced Temperature}
It is of  interest to consider another variable, 
called in \cite{Coq} the ``reduced temperature",  of the form 
\be
\tau=\frac{C_0}a,
\ee
 where $C_0>0$ is a constant, so that at the  big bang 
when $a=0$ we have  infinite ``temperature'' and the 
expansion of the universe leads to a  decay of this ``temperature''.
Strictly speaking $\tau$ is proportional to the redshift
$1+z= \frac{a(t_0)}{a(t)}$   and since the wavelength
of freely propagating particles  redshifts by this  factor
and if those  particles are in thermodynamic equilibrium  
then $\tau$ will be proportional to their temperature $T$. 
For the  standard case of radiation for which  the 
energy density is proportional to $a^{-(n +1)}$ in $n$ spatial dimensions
the distinction between the reduced temperature and the   
temperature of the fluid may be ignored.
However more generally the two are not the same.
For example in three spatial dimensions, if $P=(\gamma-1) \rho$
with $\gamma \ne \frac{4}{3}$ a constant we shall see in Subsection 3.1.1  that  
$\rho \propto T ^\frac{\gamma}{\gamma-1}$ while $\rho \propto \tau ^{3 \gamma}$.
Thus only when $\gamma=\frac{4}{3}$, i.e. for radiation,
can the two temperatures be identified.

\subsection{Conformal Time} 

It is often convenient to introduce  conformal time $\eta$ by 
\be
\dd t= a\dd\eta.
\ee
We shall  adopt  the notation $f'=\frac{\dd f}{\dd\eta}$. Then $f'=f\dot{f}$ and (\ref{a1.1}) becomes
\be
(a')^2=\frac{16\pi G_n}{n(n-1)}\rho a^4-k{a^2}.\label{a2.9}
\ee

\subsection{Chebyshev's Theorem for Binomial Integrals} 
We have seen that integration of (\ref{a1.1}) or (\ref{a2.9}) 
requires knowledge whether the binomial integral
\be\label{IaI}
I=\int x^p(\alpha+\beta x^r)^q\,\dd x,
\ee
where $p,q,r$ ($r\neq0$) are rational numbers and $\alpha,\beta$ are nonzero real numbers,
 may be represented in terms of elementary functions. As seen in \cite{CGLY}, the Chebyshev  theorem
\cite{C0,C1} is what is needed to answer such a question which states that  the integral (\ref{IaI}) is elementary
if and only if at least one of the quantities
\be\label{cd}
\frac{p+1}r,\quad q,\quad \frac{p+1}r+q,
\ee
is an integer. 

Our current work, along the line of \cite{CGLY}, is to further explore the insight that may be achieved by the Chebyshev theorem for the integration
of (\ref{a1.1}) or (\ref{a2.9}) and 
present supplementary methods when the problem is out of reach by the theorem. In particular, our main focus will be on the situations where the equation of state is nonlinear.

\section{Perfect Fluids}\label{Perfect}
\setcounter{equation}{0}

In this section we recall some standard materials about
perfect fluids as used in cosmology and their representation, when there
exits a velocity potential,  in terms
of a Lagrangian for a scalar field $\psi$ with a shift symmetry
 $\psi \rightarrow \psi + {\rm constant}$. 

In $n+1$ spacetime dimensions  the  energy-momentum tensor is  of the form
\ben
T^{\mu \nu} = \rho u^\mu u^\nu + P (g^{\mu \nu} +u^\mu u^\nu ),  
\een
where $\rho$ is energy density, $P $ is pressure and $u^\mu$ 
satisfying $g_{\mu \nu} u^\mu u ^\nu=-1$ is the (timelike)  $(n+1)$ -velocity
of the fluid. So as not to clutter up
the formula, we include any cosmological constant
in the total energy density and pressure which, in this section, we denote
by $\rho$ and $P$ with  the subscript $m$ omitted.    
For simplicity of exposition, in this section,  we shall only  consider the 
spacetime dimension  $n=3$. 
The extension to all $n$ is straightforward.

\subsection{Equations of Motion and Thermodynamics}

 Define the projection operator $h^\mu _\nu$ onto the 
orthogonal complement of the
4-velocity $u^\mu$ by
\ben
h^\mu _\nu = \delta _\nu ^\mu + u^\mu u_\nu, \qquad \Rightarrow \qquad 
h_{\mu \nu}=h_{\nu \mu},\qquad h_{\mu \nu} u^\nu =0. 
\een
Thus for any vector $V^\mu$,
\ben
V^\mu = V^\mu_{||} + V^\mu _{\perp},  
\een
with
\ben
V^\mu_{||}= - u^\mu (V _\nu u^\nu), \qquad V^\mu _{\perp} = h^\mu_\nu V^\nu,
\een
projecting the equation
\ben
\nabla _\nu T^{\mu \nu} =0, 
\een
parallel and perpendicular to $u^\mu$ results in
\bea
 (\rho + P) \nabla_\mu  u^\mu  + u^\mu \nabla _\mu \rho  &=&0,
 \label{entropy}\\
a^\mu &=& - \frac{1}{\rho + P}  h^{\mu \nu}\nabla _\nu P,   
\eea
where 
\ben
a^\mu = u^\nu \nabla _\nu u ^\mu, \qquad \Rightarrow \qquad u_\mu a^\mu =0,  
\label{Euler} \een
is the acceleration of the fluid. The \emph{expansion} $\Theta$ 
of the fluid is defined  as
\ben
\Theta = \nabla _\nu u^\nu, 
\een
and if $V$ is a comoving volume of fluid  then $\Theta = \frac{\dot V}{V}$.

We now apply the first law of thermodynamics to a volume $V$ of our fluid,
assuming no conserved charges or chemical potentials.
This states that 
\ben
\dd(\rho V)= T \dd(sV ) -P\dd V.
\een
Thus
\ben
(\dd \rho-T\dd s )V= - (\rho +P -Ts ) \dd V. 
\label{first}\een
Assuming that the fluid is homogeneous, i.e., 
that $\rho$ and $s$ do not depend upon $V$,
we deduce the \emph{Gibbs--Duhem} relation: 
\ben
\rho+P= Ts. \label{Gibbs-Duhem}
\een
The quantity $\rho+P$ is the  
\emph{enthalpy density} of the fluid and  is therefore
equal to  the product of  \emph{entropy density}
$s$  and temperature $T$. 
Now the first  law (\ref{first}) and Gibbs--Duhem relation (\ref{Gibbs-Duhem})
imply                                                                           
\ben
\dd P=s\dd T, \qquad \Leftrightarrow \qquad  \dd \rho = T\dd s,
\een
which gives in addition
\ben
\frac{\dd P}{\rho+P} = \frac{\dd T}{T}. \label{temp}
\een
Thus (\ref{entropy}) implies the \emph{conservation of entropy}  
\ben
\nabla _\mu (s u^\mu ) =0, \label{conentropy}
\een
and the \emph{Relativistic Euler equation} (\ref{Euler}) may now be rewritten as
 \ben
a^\mu = - \frac{1}{T}  h^{\mu \nu}\nabla _\nu T.  
\een

\subsubsection{Energy Conditions}

The following three energy conditions are important for cosmology:

\begin{itemize}
\item The Positive Energy  Condition: $T_{\mu \nu} V^\mu V^\mu \ge 0$ 
for every future directed timelike vector $V^\mu$. In our context this means
that $\rho\ge 0$.  Thus for a barotropic equation of state such that $P=(\gamma-1)\rho = w \rho$, where $\gamma$ is a constant, positive energy requires
$\gamma \ge  1$.

\item The Dominant Energy  Condition: $-T^\mu_\nu V^\nu$ is a future pointing
timelike or null vector.  In our context this means
that $\rho >|P|  \geq  0$. For a barotropic equation of state
with positive pressure,
dominant energy requires $\gamma \le 2$. The  limiting case
$P=\rho$ is called stiff matter and this is satisfied by  
a massless scalar field (see Subsection \ref{Lag}). For negative pressure
and positive energy, dominant energy requires $P \ge -\rho$
or $w \ge   - 1$.  
The limiting case corresponds to a positive cosmological constant     
   
\item The Strong Energy Condition: $\bigl(T_{\mu \nu}- \half g_{\mu \nu} T^\alpha _\alpha  \Bigr )  V^\mu V^\mu \ge 0$ 
for every future directed timelike vector $V^\mu$. In our context this means
that $\rho +3P \ge 0$. For a barotropic equation of state,
strong  energy requires  $\gamma \ge \frac{2}{3} $ or
$P\geq -\frac{1}{3}\rho$.   The limiting case is attained by an isotropic gas
of relativistic strings.  Incidentally an isotropic 
gas of relativistic membranes or domain walls has $ P=-\frac{2}{3} \rho$
and so violates  strong energy by a considerable margin.

\end{itemize}    

Note that for all values of $k$
\ben
\frac{\ddot a}{a}= - \frac{4 \pi G}{3} (\rho +3P)\,,
\een
and so acceleration requires a violation of the strong energy equation.
Now 
\ben
\dot H= - \frac{12 \pi G}{3} (\rho +P) +\frac{k}{a^2} 
\een  
and so if $k \le 0$, $\dot H \ge 0$ requires $\rho+P \le 0$.

\subsubsection{Examples}

If $P=(\gamma-1)\rho$ with $\gamma$  constant then
the equation of state is linear and one finds from(\ref{temp})
and the Gibbs-Duem relation (\ref{Gibbs-Duhem})  
\ben
\rho =   \bigl(\frac{T}{T_0}\bigr ) ^\frac{\gamma}{\gamma-1} \,,\qquad 
s = \gamma T_0 ^{-\frac{\gamma}{\gamma-1}} T^\frac{1}{\gamma-1}.
\een

A simple example of a nonlinear equation of state is 
provided by the Chaplygin gas  for which 
 $P=-\frac{B}{\rho}$ with $B$ constant \cite{CP,ABL,KMP}. One has 
\ben
\rho+P= \rho- \frac{B}{\rho} \,, 
\een
and therefore   if  $B< \rho^2$, the dominant 
energy condition, i.e. $\rho \ge |P|$,  holds, which
is consistent with the sound speed 
$v_s = \frac{\partial  P}{\partial \rho} = \frac{\sqrt{B } }{\rho}$ being subluminal. The strong energy condition, i.e. $\rho+3P$ holds as long as 
$3B \le \rho^2$.    
 
The temperature is given by
\bea
\frac{T}{T_0} &=& \sqrt{1- \frac{B}{\rho^2} }, \quad \rho ^2  > B,\\   
\frac{T}{T^\prime _0}
 &=& \frac{\rho ^2}{\sqrt{1- \frac{B}{\rho^2} }}, \quad \rho ^2  < B. 
\eea
 
The dependence on the scale factor is 
\ben
\rho= \sqrt{B + \frac{C}{a ^6} }, 
\een
where $C$ is a constant. If the scale factor is   large,
this behaves like
dark energy plus (if $C>0$) an admixture of stiff matter.  
At small scale factors it behaves like dust.  

There exists a generalization \cite{KMP,BS}  for which
\ben
P=  A \rho -\frac{B}{\rho ^\alpha}, 
\een 
which will be discussed in Subsection \ref{Chaplygin}.
Note that if $\alpha \ge 0$, then at large $\rho$ the effective coefficient $\gamma= A+1$
and thus   $A+1$  will determine which if any energy conditions
are satisfied  at large $\rho$.    

\subsection{Dark Energy} 

The simplest interpretation of dark energy is  
as a cosmological constant term for which
\ben
T_{\mu \nu}= - \frac{\Lambda}{8 \pi G} g_{\mu \nu}.  
\een
This has perfect fluid form with 
\ben
P_{\rm Dark} = -\rho_{\rm Dark} 
 = - \frac{\Lambda}{8 \pi G} = {\rm constant}. 
\een
Equations (\ref{entropy}) and (\ref{Euler}) are unaffected 
by the replacements
\ben
\rho \rightarrow \rho + \rho_{\rm Dark},\qquad 
P \rightarrow P -\rho_{\rm Dark},\qquad g_{\mu \nu} \rightarrow g_{\mu \nu}.
\een

\subsection{Stationary Flows}

For a \emph{stationary flow} 
\ben
u^\mu= \frac{K^\mu}{
\sqrt{-g_{\sigma v} K^\sigma K^v} }, 
\een
where   $K^\mu$ is a  timelike Killing vector field, 
one has 
\ben
a^\mu = h^{\mu \nu}  \nabla _\nu 
\ln( \sqrt{-g_{\sigma v} K^ \sigma K^v }),   
\een
whence
\ben
T \sqrt{-g_{\sigma v} K^\sigma K^v}  = {\rm constant},
\een
which is \emph{Tolman's redshifting law}.

\subsection{Irrotational   Fluids} 
The 4-velocity $u^\mu$ is vorticity free if and only if 
\ben
u \wedge \dd u =0,
\een
where $u=u_\mu \dd x^\mu$. This implies that 
\ben
u=Z^{-\half}  \dd \psi
\een
for functions $ Z $ and $\psi$. Since $u^\mu u_\mu=1$ we deduce that
\ben
Z=  -g^{\mu \nu} \partial _\mu \psi \partial _\nu \psi \,. 
\een 

\subsection{Lagrangian Description  of  Irrotational Perfect Fluids}\label{Lag}

Irrotational perfect fluids admit a Lagrangian description
in which the Lagrangian is the pressure  considered as a function
of the velocity potential $\psi$. Such media are  often referred
to  as ``$k$-essence''. The Lagrangian scalar density is given by   
\ben
{\cal L} = \sqrt{-g}  L(Z),
\een 
whence 
\ben
T_{\mu \nu} =  - \frac{2}{\sqrt{-g}}  \frac{\delta {\cal L}} 
{\delta g^{\mu \nu}} = 
+2   L_Z \partial _\mu \psi \partial _\nu \psi + g_{\mu \nu} L.   
\een
Thus
\ben\label{21}
\rho= 2ZL_z-L,\qquad P= L.
\een
Because of the shift symmetry $\psi \rightarrow \psi + {\rm constant} $,
 a conserved ``enthalpy current'' is
\ben
s^\mu = \frac{1}{\sqrt{-g}} \frac{\partial  {\cal L} }{\partial _\mu \psi}. 
\een

Some of examples are:
\begin{itemize} 
\item Stiff matter aka massless scalar $L=\half Z$, $P= \rho $. 
\item Radiation $ L= Z^2$, $P=\frac{1}{3} \rho $.  
\item {Polytrope  $L = Z^{\frac{\gamma}{2(\gamma-1)} } $,
 $P= (\gamma-1)  \rho $ }.  
\item Dark Energy $L= {\rm constant }$, $ P= - \rho$. 
\item Born-Infeld $ L=-\sqrt{1-Z} +1 $, $P= \frac{\rho}{1+\rho}$.  
\item Chaplygin $L= -\sqrt{1-Z} $, $ P= - \frac{1}{\rho}$. 
\end{itemize}
Note that the last two examples have the same equations of motion
but the energy momentum tensor differs by a dark energy term.
That is one gets to the Chaplygin case from the Born--Infeld case by
the replacements $\rho \rightarrow \rho -1$ and $P \rightarrow P+1$.  

The Polytrope equation of motion is the wave equivalent  of the
\emph{p-Laplacian}. That is 
\ben
\nabla ^\mu \bigl ( Z^{\frac{p-2}{2}} \partial _ \mu \psi  \Bigr ) =0.
\label{p-Laplace}\een
If we rewrite (\ref{p-Laplace}) as   
\ben
\partial _\mu \Bigl(\sqrt{-g} \bigl(-g^{\alpha \beta} \partial _\alpha 
\psi \partial  _\beta \psi) ^{ \frac{p-2}{2} } \bigr ) g^{\mu \nu} \partial _\nu \psi \Bigr )
=0,  
\een
with $p= \frac{\gamma}{\gamma-1}$,  we see that if $p=4$, i.e., 
for radiation, $\gamma=\frac{4}{3}$,  both the Lagrangian and the
equation of motion  are invariant under conformal rescalings
of the metric, i.e.,  under the replacement $g_{\mu \nu} \rightarrow \Omega ^2 
g_{\mu \nu} $. 

\subsection{Application to Friedmann--Lema\^{i}tre Cosmologies} 

In this case $\partial _\mu \psi= (\dot \psi,0,0,0)$ and so $Z={\dot \psi}^2$.
We regard $L(Z) $ as a function of $\dot \psi$. 

From (\ref{21})  we have 
\ben
 L(Z) = P(\dot \psi),
\een
\ben
\rho= \dot \psi \frac{\dd P}{\dd \dot \psi} -P.
\een
Now 
\ben
s^\mu = (s^0,0,0,0), 
\een
where
\ben
s^0= \frac{\dd P}{\dd \dot \psi}, 
\een
and conservation of entropy (\ref{conentropy})  gives
\ben
a^3 \frac{\dd P}{\dd \dot \psi} = {\rm constant}.
\een

Probably it is best to define say $x=\dot\psi$.

One has 
\ben
P=P(x), 
\een

\ben
\rho= x \frac{\dd P}{\dd x} -P,
\een
 \ben
a^3 \frac{\dd P}{\dd x} = c,
\een
where $c$ is a constant. For convenience, we denote $g(x)=\frac{\dd P}{\dd x}(x)$ and assume $g$ is invertible. Hence we have $x=g^{-1}(ca^{-3})$ and thus
\be 
\rho=ca^{-3} g^{-1}(ca^{-3})-\int_0^{ca^{-3}} g(\xi)\,\dd\xi-P(0).
\ee
 This equation relates the
energy density $\rho$ to the scale factor $a$ in a complicated way. 
Nevertheless, many concrete examples are manageable, with or without the Chebyshev
theorem. Below are a few of these.

  

\begin{itemize}
\item 
Set $P=\frac{x^{r+1}}{r+1}$. 
One finds 
\ben
\rho=\frac r{r+1}x^{r+1}= r P. 
\een
From \cite{CGLY} we know that the Friedmann equation is integrable if and only if $\frac1r$ is among the sequence of numbers
\be
-1,\dots,-\frac23,-\frac59,-\frac12,-\frac7{15},-\frac49,-\frac37,\dots,-\frac29,-\frac15,-\frac16,-\frac19,\frac13.
\ee
In particular, the case $r=3$, that is
\ben
ax = {\rm constant},
\een
is well known and corresponds to radiation.
\item
Chaplygin is 
\ben
L= -\sqrt{1-x^2}\,,\qquad P=-\frac{1}{\rho},
\een
and 
\ben
a ^3 \frac{x}{\sqrt{1-x^2}} =  {\rm constant}.
\een
A general form of this model will be studied in  Section \ref{Chaplygin}.
\item
Born--Infeld is
\ben
L= -\sqrt{1-x^2} +1\,,\qquad P=   \frac{\rho}{\rho+1}, 
\een
and 
\ben
a ^3 \frac{x}{\sqrt{1-x^2}} =  {\rm constant}.
\een
The general situation for  this model is thoroughly considered in Section \ref{Born}.
\item 

A model of current interest is the Bose--Einstein condensate 
considered by Chavanis \cite{Chavanis:2014hba,Chavanis:2014lra}. 

One has 
\ben
\rho= \sqrt{P_0 P} + P,
\een
where $P_0 $ is a constant.
We have 
\ben
\rho=x \frac{\dd L}{\dd x} -L,   
\een
and hence
\ben
\frac{\dd L}{2L+\sqrt{P_0 L} } = \frac{\dd x}{x}.
\een
Thus
\ben
L= \frac{1}{4} \bigl(Ax -P_0  \bigr ) ^2. 
\een
We choose the constant of integration to agree with 
the Lagrangian for stiff matter so that
\ben
P= \frac{1}{4} \bigl(\sqrt{2} x -P_0  \bigr ) ^2. 
\een 

It follows that
\ben
a^3 \sqrt{P}= {\rm constant}, 
\een
and hence
\ben
\rho= 
\frac{
\rho _{{\rm dust}\, 0} }{a^3} + 
\frac{
\rho_{ {\rm stiff matter}\, 0} 
} 
{a^6},
\een
which is a mixture of pressure matter and stiff matter
as stated in \cite{Chavanis:2014hba} and a special case of the model investigated in Section 6.

\end{itemize}

\section{The Chaplygin Gas and the Born--Infeld Models}
\setcounter{equation}{0}

In this section we study the generalized Chaplygin gas and Born--Infeld models. In the former situation, we derive exponentially expanding solutions and
obtain their exponential growth rates explicitly in terms of various physical parameters. In particular, these expressions reveal the specific roles
played by the linear terms, when nonlinear terms are present, in the equations of state. In the latter situation, we show that, although the Chebyshev
theorem is not applicable in general, an integration can still be performed to render explicit solutions.

\subsection{The Chaplygin Gas Model}\label{Chaplygin}

The equation of state of a generalized Chaplygin gas model reads 
\cite{ST,CP,ABL}
\be \label{2.1}
P_m=A\rho_m-\frac{B}{\rho^\alpha_m},
\ee
where $A,B$ are positive parameters and the exponent $\alpha$ stays in the interval $0\leq\alpha\leq1$.

Inserting (\ref{2.1}) into (\ref{1.2}) and integrating, we obtain
\be\label{2.2}
(1+A)\rho_m^{\alpha+1}=Ca^{-n(1+A)(\alpha+1)}+B,
\ee
where $C>0$ is an { integration}  constant. In fact, in our study below, the positivity condition on $A$ can be relaxed to
\be\label{xA}
A>-1.
\ee
So (\ref{xA}) will be the condition on the constant $A$ we observe in the subsequent discussion. Note that the borderline $A=-1$ is known as the Phantom Divide Line (PDL) which is an actively pursued topic \cite{V,Mc,C-L,N-P,G-C} in the research on dark matter.

In view of (\ref{8}) and (\ref{2.2}), we see that (\ref{a1.1}) takes the form
\be
\dot{a}^2=\frac{16\pi G_n}{n(n-1)}(1+A)^{-\frac1{\alpha+1}} a^2 (Ca^{-n(1+A)(\alpha+1)}+B)^{\frac1{\alpha+1}}+\frac{2\Lambda}{n(n-1)}a^2-k.
\ee

In the binomial case, $\Lambda=0,k=0$, integrability by virtue of Chebyshev's theorem is always ensured. As an illustration, we proceed to carry out the integration. 

For convenience, we
rewrite the Friedmann equation as
\be\label{2.4}
\dot{a}^2=\beta^2 a^2 (Ca^{-n(1+A)(\alpha+1)}+B)^{\frac1{\alpha+1}},\quad\beta^2=\frac{16\pi G_n}{n(n-1)}(1+A)^{-\frac1{\alpha+1}} ,
\ee
whose integration amounts to computing
\be
I=\pm\frac1\beta\int a^{-1} (Ca^{-n(1+A)(\alpha+1)}+B)^{-\frac1{2(\alpha+1)}}\,\dd a.
\ee
Since the constants $B$ and $C$ are positive, we may introduce a new variable $u$ such that
\be \label{2.6}
Ca^{-n(1+A)(\alpha+1)}+B=u^{2(\alpha+1)}.
\ee
Hence we obtain
\be\label{2.7}
I=\mp\frac2{n(1+A)\beta}\int\frac{u^{2\alpha}}{u^{2(\alpha+1)}-B}\,\dd u,
\ee
which may be integrated to yield an elementary function if and only if $\alpha$ is rational, say
\be \label{2.8}
\alpha=\frac {m_1}{m_2},\quad m_1,m_2=1,2,3,\dots.
\ee

Assuming (\ref{2.8}), we may use 
\be\label{uv}
v=B^{-\frac1{2(m_1+m_2)}} u^{\frac1{m_2}}
\ee
 to recast (\ref{2.7}) into the form
\be \label{2.10}
I=\mp \frac{2m_2}{n(1+A)\beta}B^{-\frac{m_2}{2(m_1+m_2)}}\int\frac{v^{2m_1+m_2-1}}{v^{2(m_1+m_2)}-1}\,\dd v,
\ee
and then conduct its integration.

If $\alpha$ is irrational, we may approach $\alpha$ by a sequence of rational numbers and use the above integrals to approximate the desired solution.

As an illustration, we consider the classical Chaplygin gas model \cite{KMP,BS} given simply by the equation of state 
\be\label{x511}
P_m=A\rho_m-\frac B{\rho_m},
\ee
which corresponds to $m_1=m_2=1$. Hence (\ref{2.10}) gives us the result
\bea \label{2.12}
I&=&\mp\frac2{n(1+A)\beta}B^{-\frac14}\int \frac{v^2}{v^4-1}\,\dd v\nn\\
&=&\mp\frac1{n(1+A)\beta}B^{-\frac14}\left(\frac12\ln\left|\frac{v-1}{v+1}\right|+\arctan v\right)+C_1,
\eea
where $C_1$ is an integration constant. The big bang universe $a(0)=0$ leads to the solution
\be
2n(1+A)\beta B^{\frac14}t=\ln\left(\frac{(B+Ca^{-2n(1+A)})^{\frac14}+B^{\frac14}}{(B+Ca^{-2n(1+A)})^{\frac14}-B^{\frac14}}\right)
-2\arctan\left(1+\frac{C}B a^{-2n(1+A)}\right)^{\frac14}+\pi,
\ee
which may further be resolved to yield
\be
a^{2n(1+A)}(t)=\frac C{B\left(\coth^4 \left[n(1+A)\beta B^{\frac14} t+b(t) \right]-1\right)},
\ee
where
\be\label{x5.15}
b(t)=\arctan\left(1+\frac{C}B a^{-2n(1+A)}(t)\right)^{\frac14}-\frac\pi2.
\ee
Since $b(t)$ given in (\ref{x5.15})  is a bounded function, we see that the large-time behavior of the scale factor obeys the asymptotic exponential growth law:
\be\label{x516}
a(t)\sim \left(\frac C{8B}\right)^{\frac1{2n(1+A)}}\e^{\beta B^{\frac14}t},\quad t\to\infty,
\ee
which clearly spells out the roles played by various parameters in the model since from the above the asymptotic exponential growth rate is seen to be
\be\label{x517}
\beta B^{\frac14}=4\left(\frac{\pi G_n}{n(n-1)}\right)^{\frac12} \left(\frac{B}{1+A}\right)^{\frac14}.
\ee

As another illustration, we consider the case when $\alpha=\frac12$. Thus, inserting $m_1=1,m_2=2$ into (\ref{2.10}), we have
\be \label{2.13}
I=\mp \frac{4}{n(1+A)\beta}B^{-\frac13}\int\frac{v^3}{v^6-1}\,\dd v.
\ee
To proceed further, we set $w=v^2$ in (\ref{2.13}) and use the partial fraction decomposition
\be 
\frac w{w^3-1}=\frac13\left(\frac1{w-1}-\frac{w-1}{w^2+w+1}\right),
\ee
to arrive at
\be
I=\mp \frac{1}{3n(1+A)\beta}B^{-\frac13}\left(\ln\frac{(v^2-1)^2}{v^4+v^2+1}+2\sqrt{3}\arctan\frac{2v^2+1}{\sqrt{3}}\right)+C_1,
\ee
where $C_1$ is an integration constant. As before, the big bang solution reads
\bea
3n(1+A)\beta B^{\frac13} t&=&\ln\left(\frac{\left[1+\frac CB a^{-\frac32n(1+A)}\right]^\frac23+\left[1+\frac CB a^{-\frac32n(1+A)}\right]^\frac13+1}{
\left(\left[1+\frac CB a^{-\frac32n(1+A)}\right]^\frac13-1\right)^2}\right)\nn\\
&&-2\sqrt{3}\arctan\frac1{\sqrt{3}}\left(2\left[1+\frac CB a^{-\frac32n(1+A)}\right]^\frac13+1\right)+\sqrt{3}\pi.
\eea
Resolving this relation, we obtain
\bea
&&\left(1+\frac CB a^{-\frac32n(1+A)}(t)\right)^\frac13=\nn\\
&&\frac{2+\e^{-3n(1+A)\beta B^{\frac13}t+2b(t)}+\e^{-\frac32n(1+A)\beta B^{\frac13}t+b(t)}
\left(12-3\e^{-3n(1+A)\beta B^{\frac13}t+2b(t)}\right)^{\frac12}}{2-\e^{-3n(1+A)\beta B^{\frac13}t+2b(t)}},
\eea
where 
\be
b(t)=\sqrt{3}\arctan\frac1{\sqrt{3}}\left(2 \left[1+\frac CB a^{-\frac32n(1+A)}(t)\right]^\frac13+1\right)-\frac{\sqrt{3}}2\pi
\ee
is a bounded function. Therefore we obtain the asymptotic exponential growth law as before:
\be\label{x523}
a(t)\sim \left(\frac{C}{3\sqrt{3}B}\right)^{\frac2{3n(1+A)}} \e^{\beta B^{\frac13} t},\quad t\to\infty.
\ee
Thus we also get the asymptotic exponential growth rate as follows:
\be \label{x525}
\beta B^{\frac13}=4\left(\frac{\pi G_n}{n(n-1)}\right)^{\frac12} \left(\frac{B}{1+A}\right)^{\frac13}.
\ee

For later convenience, we use (\ref{2.6}) and (\ref{uv}) to express the dependence of $a$ and $v$ directly as
\be\label{av}
v=B^{-\frac1{2(m_1+m_2)}}\left(Ca^{-n(1+A)\frac{(m_1+m_2)}{m_2}}+B\right)^{\frac1{2(m_1+m_2)}}.
\ee
Thus the big bang initial condition $a(0)=0$ with $A>-1$ leads to
\be\label{v0}
v(0)=\infty.
\ee
Hence, from (\ref{2.10}) and (\ref{v0}), we obtain the expanding solution
\be\label{tv}
\left(\frac{n(1+A)\beta}{2m_2}B^{\frac{m_2}{2(m_1+m_2)}}\right)\, t=\int_{v(t)}^\infty\frac{v^{2m_1+m_2-1}}{v^{2(m_1+m_2)}-1}\,\dd v, \quad t>0,
\ee
which monotonically climbs to the limiting level $v=1$ as $t\to\infty$.

As yet another example, we take $m_1=1$ and $m_2=3$ (i.e., $\alpha=\frac13$). Thus (\ref{tv}) yields
\bea\label{v(t)}
\frac83 n(1+A)\beta B^{\frac38}\, t&=&\ln\left(\frac{v+1}{v-1}\right)^2+4\arctan v+\sqrt{2}\ln\left(\frac{v^2-\sqrt{2} v+1}{v^2+\sqrt{2} v+1}\right)\nn\\
&&-2\sqrt{2}\left(\arctan \left[\sqrt{2} v+1\right]+\arctan\left[\sqrt{2} v-1\right]\right)\nn\\
&&+2(\sqrt{2}-1)\pi,\quad v=v(t)>1,\quad t>0.
\eea
Since all the terms except the first one on the right-hand side of (\ref{v(t)}) stay bounded, we have
\be\label{vt2}
v(t)\sim \coth\left(\frac23 n(1+A)\beta B^{\frac38}\, t\right),\quad t\to\infty.
\ee
In view of (\ref{av}) and (\ref{vt2}), we arrive at
\be
a(t)\sim \left(\frac C{16B}\right)^{\frac3{4n(1+A)}}\e^{\beta B^{\frac38}\,t},\quad t\to\infty.
\ee
Consequently, we get the asymptotic exponential growth rate as before:
\be\label{BBB}
\beta B^{\frac38}=4\left(\frac{\pi G_n}{n(n-1)}\right)^{\frac12} \left(\frac{B}{1+A}\right)^{\frac38}.
\ee

\subsection{A Universal Formula for Asymptotic Exponential Growth Rate and Cosmological Consequences}

With the explicit knowledge on exact solutions described earlier as stated in (\ref{x516}), (\ref{x523}), and (\ref{v(t)}),  we are able to derive such precise formulas 
as (\ref{x517}), (\ref{x525}), and (\ref{BBB}) for the exponential growth rate of the scale factor. In these examples, it is interesting, and perhaps surprising, to see that, in the 
context of the Chaplygin gas model, the linear part of the equation of state switched on by
the coefficient $A$ in (\ref{2.1}) also contributes to the asymptotic exponential growth rate of the scale factor, as seen in the examples (\ref{x517}), (\ref{x525}), and
(\ref{BBB}) in the previous subsection, which follow a clear pattern of the following characteristics:
\begin{itemize}
\item The growth rate contains a common factor which depends on the Newton constant $G_n$ and spatial dimensionality $n$ only.

\item The growth rate is proportional to some power $\gamma$ (say) of the common ratio $\frac B{1+A}$ composed from the linear and nonlinear coupling constants $A$ and $B$
in (\ref{2.1}). In  (\ref{x517}), (\ref{x525}), and (\ref{BBB}), we have $\gamma =\frac14,\frac13$, and $\frac38$,  corresponding to
$\alpha=1,\frac12$, and $\frac13$, in (\ref{2.1}), respectively.

\end{itemize}

Inspired by the above examples, in this subsection, we will derive a universal formula for the asymptotic exponential growth rate of the scale factor for the generalized Chaplygin model (\ref{2.1}), regardless whether $\alpha$ is rational. We then elaborate on some applications of this general formula to the study of dark matter.

For convenience, we use the new variable
\be\label{y1}
w= B^{-\frac1{2(\alpha+1)}} u
\ee
in (\ref{2.7}). Thus (\ref{2.6}) gives us
\be\label{y2}
w=\left(\frac CB a^{-n(1+A)(\alpha+1)}+1\right)^{\frac1{2(\alpha+1)}}.
\ee
Therefore the boundary condition consisting of $a(0)=0$  and  $a(\infty)=\infty$ becomes
\be\label{y3}
w(0)=\infty,\quad w(\infty)=1.
\ee
From (\ref{2.7}) and (\ref{y3}), we arrive at
\be\label{y4}
\frac12{n(1+A)\beta} B^{\frac1{2(\alpha+1)}}\, t=\int^\infty_{w(t)}\frac{w^{2\alpha}}{w^{2(\alpha+1)}-1}\,\dd w,\quad w(t)>1.
\ee
Since $w^{2(\alpha+1)}-1$ is approximated by $2(\alpha+1)(w-1)$ for $w$ near $1$, we have
\be\label{y5}
\int_{w(t)}^\infty\frac{w^{2\alpha}}{w^{2(\alpha+1)}-1}\,\dd w=\frac1{2(\alpha+1)}\int_{w(t)}^{w(t_0)}\frac1{w-1}\,\dd w+\mbox{O}(1),
\ee
where $t>t_0>0$ and $\mbox{O}(1)$ denotes a bounded quantity. Combining (\ref{y4}) and (\ref{y5}), we obtain
\be\label{y6}
w(t)=1+(w(t_0)-1)\e^{-n(1+A)(\alpha+1)\beta B^{\frac1{2(\alpha+1)}}\,t+\tiny{\mbox{O}}(1)},\quad t>t_0.
\ee
Using (\ref{y2}) and (\ref{y6}), we have the asymptotic growth pattern
\be\label{y7}
a(t)\sim C_0 \e^{\beta B^{\frac1{2(\alpha+1)}}\,t},\quad t>t_0,
\ee
where $C_0>0$ is some constant. Consequently we see that the asymptotic exponential growth rate of the scale factor $a$ is given by the {\em universal} formula
\be\label{y8}
\beta B^{\frac1{2(\alpha+1)}}=4\left(\frac{\pi G_n}{n(n-1)}\right)^{\frac12}\left(\frac B{1+A}\right)^{\frac1{2(\alpha+1)}},
\ee
which recovers the formulas (\ref{x517}), (\ref{x525}), and (\ref{BBB}) when $\alpha=1,\frac12$, and $\frac13$, respectively, and exactly confirms the
characteristics described at the beginning of this subsection.

It will be interesting to compare the above result with the
situation when the nonlinear term in the Chaplygin law is turned off, by setting $B=0$ in (\ref{2.1}). Now the conservation law  (\ref{1.2}) leads to
\be\label{y9}
\rho=\rho_0 a^{-n(1+A)},\quad A\in\bfR.
\ee
Hence, integrating the Friedmann equation (for $\Lambda=0, k=0$), we have the (well-known) expanding solution:
\be
a(t)=\left\{\begin{array}{rll} &\left(4\pi G_n\rho_0\left[\frac n{n-1}\right](1+A)^2\right)^{\frac1{n(1+A)}} t^{\frac2{n(1+A)}}, &\quad A>-1,\, t>0,\, a(0)=0;\\
& a(0)\e^{4\sqrt{\pi G_n\rho_0/n(n-1)}\,\, t},&\quad A=-1,\, t>0,\, a(0)>0;\\
&\left(a^{\frac{n(1+A)}2}(0)-4\left[\frac{\pi G_n\rho_0}{n(n-1)}\right]^{\frac12}\, t\right)^{\frac2{n(1+A)}},&\quad A<-1,\, t>0,\, a(0)>0.\end{array}\right.
\ee
So, when the nonlinear matter is absent and linear matter is that of a conventional type, $A>-1$, the scale factor of a big-bang universe fails  to grow
exponentially but instead follows a power-function type growth law, and, when $A\leq -1$, the big-bang solution with $a(0)=0$ is not allowed and the universe may only
evolve from a positive value of the scale factor, $a(0)>0$. In this latter situation, the condition $A<-1$ corresponds to a hypothetical model 
called the phantom-energy universe \cite{Hogan,Ca} where the solution blows up at finite time
\be
t_0=\frac{a^{\frac{n(1+A)}2}(0)}{4\left(\frac{\pi G_n\rho_0}{n(n-1)}\right)^{\frac12}},
\ee
known as the Big Rip \cite{Ellis,Ca2} epoch. The limiting situation, $A=-1$, is the PDL referred to earlier, where
the exponential expansion rate of the scale factor may assume arbitrarily large value, depending on the value of $\rho_0$.

On the other hand, the  universal formula (\ref{y8}) indicates that, within the context of the generalized Chaplygin fluid (\ref{2.1}),
a conventional ($A>-1$) linear term alone does not give rise to exponential growth but 
it plays a role
in exponential growth when a nonlinear term is present, no matter how weak this nonlinear matter is, and that,
this {\em linear} contribution becomes significant near the PDL limit. That is,  the asymptotic exponential expansion rate of the scale factor may assume arbitrarily large values for $A>-1$ and $A$ stays near $-1$. In other words, a tiny presence of the Chaplygin matter would switch on the presence of dark matter
which becomes significant near the PDL limit.

\subsection{The Born--Infeld Type Fluid Model}\label{Born}


In this subsection, we shall  consider a Born--Infeld type 
fluid whose equation of state is of the form
\be\label{5.1}
P_m=\frac{\rho_m}{\rho_m+1}.
\ee
As we saw in Section \ref{Perfect}, there is a close relation
between  the Born--Infeld
and  the classical Chaplygin gas models. The two Lagrangians differ by a constant  
which adds a cosmological or dark energy term.   Thus if 
$\tilde \rho_m= \rho_m  +1$ and $\tilde P_m= P_m-1$ we find
\ben
\tilde P_m= - \frac{1}{\tilde \rho_m}   
\een
However, the Born--Infeld equation of state, unlike the Chaplygin gas,
 satisfies
 both the dominant and strong energy conditions
for all values of the energy density.
 
Inserting (\ref{5.1}) into (\ref{1.2}) gives us the following dependence relation between the mass density and the scale factor:
\be\label{5.2}
\rho_m=-1+\sqrt{1+C_0 a^{-2n}},
\ee
where $C_0>0$ is a constant.
Hence, in view of (\ref{8}),  the Friedmann equation (\ref{a1.1}) becomes
\be\label{5.3}
H^2=\frac{16\pi G_n}{n(n-1)}\left(-1+\sqrt{1+C_0 a^{-2n}}+\frac\Lambda{8\pi G_n}\right)-\frac k{a^2}.
\ee

First, we see that, in order to be able to utilize the Chebyshev theorem for integration of (\ref{5.3}), we have to assume $k=0$ and
\be\label{5.4}
\Lambda=8\pi G_n.
\ee
In such a situation, the Friedmann equation (\ref{5.3}) is reduced into (\ref{2.4}) which corresponds to the Chaplygin gas model
(\ref{2.1}) with $\alpha=1,A=0,B=1$ and whose solution is given
in (\ref{2.12}) with $C=C_0$ in (\ref{2.6}).
Thus various solutions of specific interests may be written down immediately. For example, the solution satisfying $a(0)=0$ is given by
\be
\ln\left|\frac{u-1}{u+1}\right|+2\arctan u=\pi \mp 8\sqrt{\frac{n\pi G_n}{n-1}}\, t,\quad u=\left(1+C_0 a^{-2n}\right)^{\frac14}.
\ee

Next, we note that, when (\ref{5.4}) is not fulfilled and, so, the Chebyshev theorem fails to be directly applicable, we can still integrate (\ref{5.3}) at $k=0$. For this purpose, we set
\be
\alpha=\frac{\Lambda}{8\pi G_n}-1,\quad \beta^2=\frac{16\pi G_n}{n(n-1)},
\ee
and rewrite (\ref{5.3}) as
\be\label{5.7}
(\dot{a})^2=\beta^2 a^2 \left(\alpha+\sqrt{1+C_0 a^{-2n}}\right),
\ee
whose integration reads
\be\label{5.8}
I=\int a^{-1}\left(\alpha+\sqrt{1+C_0a^{-2n}}\right)^{-\frac12}\,\dd a=\pm\beta t.
\ee
Obviously we cannot apply the Chebyshev theorem as before. Nevertheless, we may set $1+C_0 a^{-2n}=u^2$ so that $a=C_0^{\frac1{2n}}(u^2-1)^{-\frac1{2n}}$ to convert
the left-hand side of (\ref{5.8}) into
\be\label{5.9}
I=-\frac1n\int u(u^2-1)^{-1} (\alpha+u)^{-\frac12}\,\dd u.
\ee
Now set $\alpha+u=w^2$ or $u=w^2-\alpha$. Then (\ref{5.9}) becomes
\bea\label{5.10}
I&=&-\frac2n\int\frac{w^2-\alpha}{(w^2-\alpha)^2-1}\,\dd w\nn\\
&=&-\frac1n\int\left(\frac1{w^2-(\alpha+1)}+\frac1{w^2-(\alpha-1)}\right)\,\dd w,
\eea
whose integration up to an additive integration constant is readily computed to give us the results:
\be
I=\left\{\begin{array}{lll}-\frac1n\left(\frac1{2\sqrt{1+\alpha}}\ln\left|\frac{w-\sqrt{1+\alpha}}{w+\sqrt{1+\alpha}}\right|+\frac1{\sqrt{1-\alpha}}\arctan\frac w{\sqrt{1-\alpha}}\right),&\quad
|\alpha|<1,\\
-\frac1n\left(\frac1{2\sqrt{2}}\ln\left|\frac{w-\sqrt{2}}{w+\sqrt{2}}\right|-\frac1w\right),&\quad \alpha=1,\\
-\frac1{2n}\left(\frac1{\sqrt{1+\alpha}}\ln\left|\frac{w-\sqrt{1+\alpha}}{w+\sqrt{1+\alpha}}\right|+\frac1{\sqrt{\alpha-1}}\ln\left|\frac{w-\sqrt{\alpha-1}}{w+\sqrt{\alpha-1}}\right|\right),&\quad \alpha>1,\\
-\frac1n\left(\frac1{\sqrt{2}}\arctan\frac w{\sqrt{2}}-\frac1w\right),&\quad \alpha=-1,\\
-\frac1n\left(\frac1{\sqrt{-(\alpha+1)}}\arctan\frac w{\sqrt{-(\alpha+1)}}+\frac1{\sqrt{1-\alpha}}\arctan\frac w{\sqrt{1-\alpha}}\right),&\quad \alpha<-1.
\end{array}\right.
\ee

Thus, in short, we conclude that the Born--Infeld fluid model (\ref{5.1}) allows an exact integration when $k=0$ for any value of the cosmological constant, although it lies beyond the
scope of the Chebyshev theorem.

To end this section, we examine the case $\alpha>1$ closely as an illustrative example. It is readily seen that the big bang solution, with $a(0)=0$ and
$a(t)>0$ for $t>0$, is allowed which is given by
\bea
2n\beta t&=&\frac1{\sqrt{1+\alpha}}\ln\left|\frac{\sqrt{\alpha+\sqrt{1+C_0 a^{-2n}}}+\sqrt{1+\alpha}}{\sqrt{\alpha+\sqrt{1+C_0 a^{-2n}}}-\sqrt{1+\alpha}}\right|\nn\\
&&+\frac1{\sqrt{\alpha-1}}\ln\left|\frac{\sqrt{\alpha+\sqrt{1+C_0 a^{-2n}}}+\sqrt{\alpha-1}}{\sqrt{\alpha+\sqrt{1+C_0 a^{-2n}}}-\sqrt{\alpha-1}}\right|.
\eea
Thus, in the limit $a(t)\to\infty$ as $t\to\infty$, since the second term on the right-hand side of the above is bounded, we have the asymptotic formula:
\bea
a(t)&\sim&\left(\frac{C_0}{8(1+\alpha)}\right)^{\frac1{2n}}\e^{\beta\sqrt{1+\alpha}\, t}\nn\\
&=&\left(\frac{\pi G_n C_0}{\Lambda}\right)^{\frac1{2n}}\e^{\sqrt{\frac{2\Lambda}{n(n-1)}}\, t},\quad t\to\infty.
\eea
It is interesting to note that the Newton constant is absent from the exponent of the asymptotic growth law.

\section{Multiple Fluids}\label{multiple}
\setcounter{equation}{0}

In this section we study the integrability problem for several multiple fluid models in view of the Chebyshev theorem. In a two-term energy density model, we give
an example of explicit solution that obeys a linear asymptotic growth law. In a two-fluid model, we show how to use the information from integration to obtain
periodic solutions linking big-bang to big crunch epochs. In a trinomial Friedmann equation model, we find exponentially expanding solutions.

\subsection{A Two-Term Energy Density Model}\label{Two} 

Consider now the classical Friedmann equation in three-spatial dimensions so that (\ref{a1.1}) assumes the form
\be\label{3.1}
H^2=\frac{8\pi G}3\rho-\frac k{a^2}.
\ee
In \cite{NP1,NP2} the energy density is taken to enjoy the general decomposition
\be\label{3.2}
\rho=\sum_{n=-\infty}^\infty \rho^+_n a^{-n}+\sum_{m=-\infty}^\infty \rho_m^- a^{-m},
\ee
where the quantities $\rho_n^+$'s are positive constants and, $\rho^-_m$'s, negative. Following \cite{NP1,NP2}, we study the simple situation that only two terms in (\ref{3.2}) are present
and $k$ is set to zero so that the Friedmann equation becomes
\be \label{3.3}
\frac{\dot{a}^2}{H^2_0}=\Om_n^+ a^{2-n}+\Om_m^- a^{2-m},
\ee
where $\Om^+_n>0$ and $\Om^-_m<0$ are the normalized positive and negative energy density coefficients and $H_0>0$ is the Hubble parameter at a time when $a=1$, although our study here does not need to be confined within such a sign condition regarding the quantities $\Om^+_n$ and $\Om^-_m$. Hence, from
(\ref{3.3}), we arrive at the integral
\be\label{3.4}
I=\frac1{H_0}\int a^{\frac n2 -1}(\Om^+_n+\Om^-_m a^{n-m})^{-\frac12}\,\dd a =\pm t.
\ee
When $m=n$, (\ref{3.4}) is trivially computed. Thus, in the sequel we assume $m\neq n$. In view of the Chebyshev theorem we see that
(\ref{3.4}) may be integrated if and only if $m,n$ ($m\neq n$) satisfy one of  the following conditions:
\bea
n&=&\frac{2N}{2N-1}m,\quad N=0,\pm1,\pm2,\dots,\label{3.5}\\
m&=&\frac{2N}{2N+1}n,\quad N=0,\pm1,\pm2,\dots.
\eea
In particular, if $m=0$ or $n=0$, integrability is ensured for any integer $n$ or $m$.

As an illustration, we show how to integrate (\ref{3.4}) under the condition (\ref{3.5}). In fact, in this case, (\ref{3.5}) may be rewritten as
\be \label{3.7}
I=\frac1{H_0}\int a^{\frac{mN}{2N-1}-1}\left(\Om^+_n+\Om^-_m a^{\frac m{2N-1}}\right)^{-\frac12}\,\dd a.
\ee
Now set
\be\label{x58}
\Om^+_n+\Om^-_m a^{\frac m{2N-1}}=u^2,
\ee
which leads to the inverse relation
\be\label{59}
a=\left(\frac{u^2}{\Om_m^-}-\frac{\Om^+_n}{\Om_m^-}\right)^{\frac{2N-1}m},
\ee
so that (\ref{3.7}) reduces into the integral
\be\label{x510}
I=\frac{2(2N-1)}{m H_0 (\Om^-_m)^N}\int (u^2-\Om^+_n)^{N-1}\,\dd u,
\ee
that allows a recursive computation with the substitution $u=\sqrt{\Om^+_n}\sec\theta$, say. Note that, from (\ref{x58}), we have the restriction
\be\label{xx511}
\Om_n^+-u^2=|\Om_m^-| a^{\frac{m}{2N-1}}\geq 0
\ee
for the range of the variable $u$. 

After the above general discussion, we consider a concrete situation with $m=3,n=2, N=-1$. Thus, from integrating (\ref{x510}), we get
\be\label{512}
\pm\frac{H_0}{|\Om_3^-|}t=-\frac{u}{\Om_2^+(u^2-\Om_2^+)}+\left(\Om_2^+\right)^{-\frac32}\arctanh\left(\frac{u}{\sqrt{\Om_2^+}}\right)+C,
\ee
where $C$ is an integration constant. On the other hand, since $N=-1$, we need to strengthen (\ref{xx511}) into
\be\label{513}
\Om_2^+-u^2=|\Om_3^-|a^{-1}>0.
\ee
In particular, we have the lower bound
\be\label{514}
a(0)\geq \frac{|\Om_3^-|}{\Om_2^+}
\ee
for the initial value of the scale factor so that a big-bang universe is not allowed. Fix the constant $C$ in (\ref{512}) with the initial value of the scale factor at
$t=0$ satisfying (\ref{514}). Choose the plus sign in (\ref{512}) representing an expanding universe. Then, from (\ref{512}), we see that
\be\label{515}
\lim_{t\to\infty} u(t)=\sqrt{\Om_2^+},
\ee
or $a(t)\to\infty$ as $t\to\infty$. Since the first term in (\ref{512}) is dominant near $u=\sqrt{\Om_2^+}$, we arrive, in view of (\ref{513}), at the asymptotic 
linear growth law:
\be
a(t)\sim \sqrt{\Om_2^+}\, H_0\,t,\quad t\to\infty.
\ee
It is interesting, and perhaps surprising, to note that $\Om_3^-$ makes no appearance here.

Another interesting example that fits our study is the Friedmann equation
\be\label{xH}
H^2=\frac{8\pi G}3 (\rho_d+\rho_b+\rho_\gamma),
\ee 
considered in \cite{Aviles} in the context of the so-called geometrothermodynamics (GTD) \cite{Q}, where $\rho_d$ is the GTD dark matter density given
in terms of the scale factor $a$ by
\be
\rho_d=\rho_{d0}\left({\cal A}a^{-3(\alpha-\beta)}+(1-{\cal A}) a^{-3(1+\alpha)}\right)^{\frac1{1+\alpha}},\quad {\cal A}>0,\alpha,\beta\in\bfR,
\ee
and $\rho_b=\rho_{b0}a^{-3}$ and $\rho_\gamma=\rho_{\gamma0} a^{-4}$ are the baryonic and relativistic matter densities. Thus, when some terms in
these densities are absent so that the integration of (\ref{xH})  is of a binomial type, the Chebyshev theorem becomes applicable. Here we discuss such a situation.

For simplicity, we assume the baryonic and relativistic matter densities in (\ref{xH}) are absent. Hence we arrive at the integration
\be
\pm 2\sqrt{\frac{2\pi G}3}\, t=I_{\alpha,\beta}\equiv \int a^{\frac12}\left((1-{\cal A})+{\cal A} a^{3(1+\beta)}\right)^{-\frac1{2(1+\alpha)}}\,\dd a.
\ee
Thus, in view of the Chebyshev theorem, we see that, when $\alpha, \beta$ are rational numbers, $I_{\alpha,\beta}$ is elementary if and only if one of the following three quantities
\be
\frac1{2(1+\alpha)},\quad \frac1{2(1+\beta)},\quad \frac1{2(1+\alpha)}- \frac1{2(1+\beta)}=\frac{\beta-\alpha}{2(1+\alpha)(1+\beta)}
\ee
is an integer. For example, when $\alpha=\beta=-\frac32$, we have
\be
I_{-\frac32,-\frac32}={\cal A} \ln a+\frac23(1-{\cal A}) a^{\frac32}+C,
\ee
where $C$ is an integration constant. Hence we see that the big bang solution, $a(0)=0$, is not allowed, and when ${\cal A}<1$ the scale factor obeys a power
law growth pattern and when ${\cal A}=1$ an exponential pattern.

\subsection{A Two-Fluid Model} \label{Twofluid}

Consider the equation of state of a two-fluid model
\be\label{6.1}
\rho_m=\frac{2R}{a^3}+\frac{A^2-R^2}{a^4},
\ee
where $A,R>0$ are constants and both 
cases $A>R$ and $A<R$ are of cosmological interest.
If $n=3$, the first term corresponds to pressure free matter
and the second to radiation.  
We are interested in a closed universe with vanishing cosmological constant. 
In that case   the Friedmann equation reads
\be\label{6.2}
(\dot{a})^2=\frac{16\pi G_n}{n(n-1)}\,\frac{\left(A^2-R^2+2Ra -B_n a^2\right)}{a^2},
\ee
where
\be\label{6.3}
B_n=\frac{n(n-1)}{16\pi G_n}.
\ee
The necessary condition for (\ref{6.2}) to be meaningful, that is, the right-hand side of (\ref{6.2}) is positive for some values of $a$, is
\be\label{6.4}
R^2<A^2+\frac{R^2}{B_n},
\ee
which guarantees that the quadratic function
\be
q(a)=A^2-R^2+2Ra -B_n a^2
\ee
has two real roots, which are
\be\label{xrts}
a_{\pm}=\frac{R}{B_n}\pm\sqrt{\left(\frac R{B_n}\right)^2+\frac1{B_n}(A^2-R^2)}.
\ee

If $A>R$,  then $a_+>0$ and $a_-<0$; if $A\leq R$, then $a_+>a_-\geq0$. In either case, we may rewrite (\ref{6.2}) as
\be\label{97}
(\dot{a})^2=\frac{(a-a_-)(a_+-a)}{a^2},
\ee
where $a$ stays in the interval $0\leq a\leq a_+$ if $A>R$; in $a_-\leq a\leq a_+$ if $A\leq R$, which formally leads to its integration
\be\label{6.8}
I=\int a\left((a-a_-)(a_+-a)\right)^{-\frac12}\,\dd a=\pm  t.
\ee
Setting $u=a-a_-$, we have
\bea
\left(\sqrt{\left(\frac R{B_n}\right)^2+\frac1{B_n}(A^2-R^2)}-\frac{R}{B_n}\right)
&=&-a_-\leq u\leq a_+-a_-\nn\\
&=&2\sqrt{\left(\frac R{B_n}\right)^2+\frac1{B_n}(A^2-R^2)},
\eea
and we can rewrite the integral $I$ on the left-hand side of (\ref{6.8}) as
\be
I=\int u^{\frac12}\left((a_+-a_-)-u\right)^{-\frac12}\dd u +a_-\int u^{-\frac12}\left((a_+-a_-)-u\right)^{-\frac12}\dd u\equiv I_1+I_2.
\ee
It is easily seen that the Chebyshev theorem implies that both $I_1$ and $I_2$ are integrable to yield elementary functions. Carrying out the integrations, we obtain
\be \label{x521}
I=\frac12(a_++a_-)\arctan\left(\frac{(a-a_+)+(a-a_-)}{2\sqrt{(a_+-a)(a-a_-)}}\right)-\sqrt{(a_+-a)(a-a_-)}+C,
\ee 
where $C$ is an integration constant. If $A>R$, the big bang solution, $a(0)=0$, is allowed; if $A\leq R$, the initial value of the scale factor has to satisfy
$a(0)\geq a_-$.  These  fix the constant $C$. 

Below we consider the big bang solution with the assumption $A>R$. Thus $C$ in (\ref{x521})  is given as
\be\label{xx522}
C\equiv C_0=\sqrt{a_+(-a_-)}+\frac12(a_++a_-)\arctan\left(\frac{a_++a_-}{2\sqrt{a_+(-a_-)}}\right).
\ee
It is clear that the solution
increases monotonically until it hits $a=a_+$ when $t$ becomes 
\be\label{xx523}
t=t_s=C_0+\frac\pi4(a_++a_-).
\ee
 From (\ref{97}) we find $\dot{a}(t_s)=0$. Furthermore, differentiating (\ref{97}) and taking the limit $t\to t_s^-$, we have
\be
\ddot{a}(t_s)=-\frac{(a_+-a_-)}{2a^2_+}<0.
\ee
This indicates that
 the solution starts to decrease beyond $t_s$. In fact, if we use $a(t)$ to denote the solution obtained above in the time interval $0\leq t\leq t_s$, the solution
of (\ref{97}) that continues to evolve beyond $t_s$ is given by
\be
a(2t_s-t),\quad t_s\leq t\leq 2t_s,
\ee
as can be verified directly. At the time $2t_s$, the solution vanishes again and then repeats the same cycle. That is, we have obtained a periodic solution which starts from $a=0$ at $t=0$, climbs to
its maximum height $a_+$ at half period $\frac12 t_p=t_s$, and then starts to descend steadily until $a=0$ at a full period $t_p=2t_s$.
The quantity $t_s$ may be called  \emph{time of  maximum expansion}. In fact 
$t_s$ is a \emph{moment of time symmetry}
since  the solution curve of the scale factor against time  is
clearly symmetric about its time of maximum expansion.  
Typically such solutions expand from a ``Big-Bang'' and recollapse to a
``Big Crunch''. In the case of a negative cosmological constant, 
the Big-Bang is a coordinate singularity and the 
solution may be extended to a  periodic solution.  In the case of 
a barotropic equation of state, both the Big Bang and the Big Crunch
are singular there is no unambiguous periodic continuation.  

For the record, we can explicitly rewrite in view of (\ref{xrts}), (\ref{xx522}), and (\ref{xx523}) the period $t_p$ just obtained as follows:
\bea
t_p&=&2C_0+\frac\pi2(a_++a_-)\nn\\
&=&8\left(\sqrt{\frac{\pi G_n(A^2-R^2)}{n(n-1)}}+\frac{2\pi^2 G_n R}{n(n-1)}+\frac{4\pi G_n R}{n(n-1)}\arctan\left[\frac{4R\sqrt{\pi G_n}}{\sqrt{n(n-1)(A^2-R^2)}}\right]\right).\nn\\
\eea

Thus we have arrived at a comprehensive and accurate understanding of 
the evolution of the scale factor in the two-fluid closed-universe model with respect to
cosmic time.
\medskip

The problem of this section motivates us to study the equation of state for the situation that the mass density and scale factor are related simply by
\be \label{813}
\rho_m=\frac A{a^r}+\frac B{a^s},\quad r\neq s,
\ee
where $A,B,r,s$ are constants. Using (\ref{813}) in the conservation law (\ref{1.2}) which may be rewritten in the form
\be
P_m=-\rho_m -\frac an\frac{\dd\rho_m}{\dd a},
\ee
we find
\be \label{815}
P_m=\frac A{a^r}\left(\frac rn-1\right)+\frac B{a^s}\left(\frac sn-1\right).
\ee
We are interested in the case when $P_m$ is simply a power function of $a$. Thus, without loss of generality, we may assume
\be 
r=n,
\ee
which leads us to arrive at 
\be 
a^s P_m=B\left(\frac sn -1\right),\quad s\neq n,
\ee
and the equation of state
\be
\rho_m=\frac{n P_m}{(s-n)}+A\left(\frac{nP_m}{B(s-n)}\right)^{\frac ns}.
\ee




\newcommand{\T}{\tau}

\subsection{A Trinomial Friedmann Equation}\label{Trinomial}

We consider the trinomial Friedmann equation with $k=1$ and a nonzero cosmological constant $\Lambda$ of the form \cite{AC}
\be 
H^2=\frac{16\pi G_n}{n(n-1)}\rho_m +\frac{2\Lambda}{n(n-1)}-\frac1{a^2},
\ee
where the mass density $\rho_m$ is taken to obey the specific relation 
\be
\rho_m=\frac{C_1}{a^{n+1}},
\ee
in which $C_1$ is a constant, so that an integration of the equation leads to
\be
\pm t=I_n=\int\left(-1+\frac{2\Lambda}{n(n-1)} a^2+\frac{16\pi G_n C_1}{n(n-1)}a^{-(n-1)}\right)^{-\frac12}\,\dd a.
\ee
For $n=3$ and $\Lambda>0$ (say),  we have
\bea \label{4.4}
I_3&=&\int a\left(-a^2+\frac{\Lambda}3 a^4+\frac{8\pi G_3C_1}3\right)^{-\frac12}\,\dd a\nn\\
(\mbox{setting } u=a^2)&=&\frac12\sqrt{\frac3\Lambda}\int\left( u^2-\frac3\Lambda u+\frac{8\pi G_3C_1}\Lambda\right)^{-\frac12}\,\dd u,
\eea
whose integration is studied in \cite{AC} using the Weierstrass elliptic functions. Here we attempt a direct integration in view of Chebyshev's theorem and in terms of elementary functions.
For this purpose, we assume
\be\label{4.5}
32\pi G_3C_1\Lambda<9,
\ee
(the critical case when $32\pi G_3 C_1=9$ is trivial) which allows us to rewrite (\ref{4.4}) as
\be 
I_3=\frac12\sqrt{\frac3\Lambda}\int\left((u-u_1)(u-u_2)\right)^{-\frac12}\,\dd u,
\ee
where
\be
u_{1,2}=\frac3{2\Lambda}\pm\frac12\sqrt{\left(\frac3\Lambda\right)^2-\frac{32\pi G_3C_1}\Lambda}
\ee
are ensured to be real and $u_1>u_2$ in view of (\ref{4.5}). Setting $v=u-u_1$ (say), we get
\be \label{x56}
I_3=\frac12\sqrt{\frac3\Lambda}\int v^{-\frac12}(v+(u_1-u_2))^{-\frac12}\,\dd v,
\ee
which is integrable by the Chebyshev theorem.  Hence, carrying out the integration (\ref{x56}), we find
\bea\label{I3}
I_3&=&\frac12\sqrt{\frac3\Lambda}\ln\left(-\frac12(u_1+u_2)+u+\sqrt{u^2-(u_1+u_2)u+u_1u_2}\right)+C\nn\\
&=&\frac12\sqrt{\frac3\Lambda}\ln\left(-\frac3{2\Lambda}+u+\sqrt{u^2-\frac3{\Lambda}u+\frac{8\pi G_3 C_1}{\Lambda}}\right)+C\nn\\
&=&\frac12\sqrt{\frac3\Lambda}\ln\left(-\frac3{2\Lambda}+a^2+\sqrt{a^4-\frac3{\Lambda}a^2+\frac{8\pi G_3 C_1}{\Lambda}}\right)+C,
\eea
where $C$ is an integration constant. 

It is readily checked that the condition (\ref{4.5}) does not permit a big bang solution with $a(0)=0$. In fact, to stay in the real regime, it 
suffices to impose the condition
\be
a(0)\geq\sqrt{ u_1}=\left(\frac3{2\Lambda}+\left[\left(\frac3{2\Lambda}\right)^2-\frac{8\pi G_3 C_1}{\Lambda}\right]^{\frac12}\right)^{\frac12}
\ee
for the initial value of the scale factor. Hence the expanding-universe solution following from (\ref{I3}) is given by the formula
\be
a^2+\sqrt{a^4-\frac3{\Lambda}a^2+\frac{8\pi G_3 C_1}{\Lambda}}=\frac3{2\Lambda}+C_0 \e^{2\sqrt{\frac{\Lambda}3}t},
\ee
where $C_0>0$ is a constant depending on $a(0)$. Consequently, we have the following exponential growth pattern for the asymptotic form of the solution:
\be
a(t)\sim \sqrt{\frac{C_0}2}\e^{\sqrt{\frac\Lambda3}t},\quad t\to\infty,
\ee
which spells out the role of the cosmological constant $\Lambda$ clearly.

On the other hand, it can be examined that $I_n$,  when $n>3$, does not allow a reduction into an integration involving a binomial differential so that the Chebyshev theorem is applicable.

\subsection{Friedmann's Equation in a Chern--Simons Modified Gravity Theory}

Following Jackiw and Pi \cite{Jackiw}, the Chern--Simons modified gravity theory in four spacetime dimensions is based on the extended Hilbert--Einstein action
\be
S=\frac1{16\pi G}\int \left\{\sqrt{-g} R+\frac\theta4 \, ^*RR\right\} \dd^4 x,
\ee
where $R$ is the Ricci scalar,  $^* RR$ the Pontryagin term given by
\be
^*RR= \,^* R_{\beta}^{\alpha\mu\nu}R^{\beta}_{\alpha\mu\nu},
\ee
with $R^{\beta}_{\alpha\mu\nu}$ the Riemann tensor and $^*R_{\beta}^{\alpha\mu\nu}= \frac12\epsilon^{\mu\nu\mu'\nu'}R^\alpha_{\beta\mu'\nu'}$ the dual 
Riemann tensor,  $\theta$ a prescribed quantity, and the cosmological constant is taken to be zero. Recall that the Pontryagin term $^*RR$ may be represented as a total divergence \cite{Jackiw,Gru,Fur}:
\be
\frac14 {^*R R}=\nabla_\mu K^\mu,
\ee
where the 4-current $K^\mu$ reads
\be
K^\mu=\epsilon^{\mu\nu\alpha\beta}\left(\Gamma^{\nu'}_{\nu\mu'}\pa_\alpha\Gamma^{\mu'}_{\beta\nu'}+\frac23 \Gamma^{\nu'}_{\nu\mu'}
\Gamma^{\mu'}_{\alpha\alpha'}\Gamma^{\alpha'}_{\beta\nu'}\right),
\ee
in terms of the Christoffel symbols, resembling a Chern--Simons invariant \cite{CS1,CS2}.
 
In \cite{Fur2},  a general Chern--Simons modified gravity model defined by the action
\be
S=\frac1{16\pi G}\int \left\{\sqrt{-g} R+\frac l4\Theta\, ^*RR-\frac12\pa^\mu\Theta\pa_\mu\Theta\right\}\dd^4 x +S_{\mbox{\small mat}},
\ee
is considered, where $\Theta$ is now a dynamical variable \cite{Sm}, of the flavor of the Bekenstein model \cite{Bk0,Bk1,Bk2}, $l$ a coupling parameter, and $S_{\mbox{\small mat}}$ denotes an additional matter action contribution.
In the Friedmann--Robertson--Walker metric limit governing a homogeneous and isotropic universe, the scalar field $\Theta$ is shown \cite{Fur2} to obey the
dynamics
\be
\ddot{\Theta}+3 H\dot{\Theta}=0,
\ee which gives rise to the first integral
\be
\dot{\Theta}=Ca^{-3},
\ee
where $C$ is a constant, and the energy conservation law reads
\be 
\frac{\dd}{\dd t}\left(\rho+\frac{C^2}2 a^{-6}\right)+3H(\rho+P+C^2 a^{-6})=0,
\ee
which is seen to simplify into the usual one
\be \label{564}
\dot{\rho}+3H(\rho+P)=0,
\ee
independent of the Chern--Simons dynamical constant $C$ since $H=\frac{\dot{a}}a$. On the other hand,
the flat-space ($k=0$) Friedmann equation is
\be
H^2=\frac{8\pi G}3\left(\rho+\frac{C^2}2 a^{-6}\right),
\ee
which relies on $C$. Thus, with the linear equation of state 
\be
P=(\gamma-1)\rho,
\ee
we read off from (\ref{564}) the relation $\rho=\rho_0 a^{-3\gamma}$. Consequently we arrive at the modified Friedmann equation \cite{Fur2}:
\be\label{567}
\left(\frac{\dot a}a\right)^2=\frac{8\pi G}3\left(\rho_0 a^{-3\gamma}+\frac{C^2}2 a^{-6}\right)\equiv \frac{8\pi G}3\rho_{\mbox{\small eff}},
\ee
where $\rho_{\mbox{\small eff}}$ is the effective energy density given by
\be
\rho_{\mbox{\small eff}}=\rho_{\mbox{\small eff}}(a)=\rho_0 a^{-3\gamma}+\frac{C^2}2 a^{-6},
\ee
resembling another two-fluid model.

Integrating (\ref{567}), we find
\be
I=\int a^2\left(\rho_0 a^{6-3\gamma}+\frac{C^2}2\right)^{-\frac12}\dd a=\pm 2\sqrt{\frac{2\pi G}3}\, t.
\ee
In view of the Chebyshev theorem, we see that, when $\gamma$ is rational, the integral $I$ is elementary if and only if $\gamma$ assumes the values
\bea
\gamma&=&2-\frac1N,\quad N=\pm1,\pm2,\dots,\label{570}\\
\gamma&=&2-\frac2{2N+1},\quad N=0,\pm1,\pm2,\dots.\label{571}
\eea

As some illustrations, we first choose $N$ among the sequence (\ref{570}), say $N=1$ or $\gamma=1$ (dust), to get
\be
I=\frac2{3\rho_0}\left(\rho_0 a^3+\frac{C^2}2\right)^{\frac12}+C_1,
\ee
where $C_1$ is an integration constant. Hence the big-bang initial condition $a(0)=0$ gives us the expanding solution
\be
a(t)=\left(6\pi G\rho_0  t^2+2\sqrt{3\pi G}|C|t\right)^{\frac13},\quad t\geq0.
\ee
Similarly we choose $\gamma$ among the second sequence (\ref{571}), say $N=1$ or $\gamma=\frac43$ (radiation), to get
\be\label{574}
I=\frac{a}{2\rho_0}\left(\rho_0 a^2+\frac{C^2}2\right)^{\frac12}-\frac{C^2}{4\rho_0^{\frac32}}
\ln\left(\sqrt{\rho_0} \, a+\left(\rho_0 a^2+\frac{C^2}2\right)^{\frac12}\right)+C_1,
\ee
where $C_1$ is an integration constant so that for the big-bang solution it is fixed to be
\be
C_1=\frac{C^2}{4\rho_0^{\frac32}}\ln\frac{|C|}{\sqrt{2}}.
\ee
From (\ref{574}), we obtain the following asymptotic behavior for expanding solutions:
\be
a(t)\sim 2\left(\frac{2\pi G\rho_0}3\right)^{\frac14} t^{\frac12},\quad t\to\infty.
\ee

So in both cases the solutions increase as  power functions.

\medskip

Thus, we have seen that, in the  multiple fluid situations studied above,  power-law growth, periodic oscillations, and exponential expansion may all occur.
It is interesting to note that a very wide range of evolutionary behavior can be achieved in the context of 
appropriate single-fluid models \cite{JB1} as well. See also \cite{JB2}.

\section{The Reduced Temperature Equation}
\label{Reduced}
\setcounter{equation}{0}

In this section we study the  Friedmann equation  in terms of the reduced temperature and  conformal time. We
will see that, in such
a setting,  although the Chebyshev theorem is no longer applicable, new insight regarding integrability may be acquired which complements what may be obtained in terms of the scale factor and cosmic time. 

\subsection{Governing Equation in Conformal Time}

We follow \cite{Coq} to consider the normalized Friedmann equation
\be\label{10.1}
\frac1{a^2}\left(\frac{\dd a}{\dd t}\right)^2=\frac{C_r}{a^4}+\frac{C_m}{a^3}-\frac{k}{a^2}+\frac\Lambda3,
\ee
in 3 dimensions, where $C_r$ and $C_m$ are radiation and matter parameters, respectively. In the zero cosmological constant limit, $\Lambda=0$, the equation
can be integrated in view of the discussion in Section \ref{multiple}. In this section, we study what happens when we switch to the conformal time $\eta$:
\be
\dd t=a\dd \eta,
\ee
and reformulate the Friedmann equation in the reduced temperature
\be
\T=\frac1{\sqrt{\Lambda_c}\,a},\quad\sqrt{\Lambda_c}=\frac2{3C_m},
\ee
as in \cite{Coq}. In terms of these new
variables, the Friedmann equation (\ref{10.1}) becomes \cite{Coq}
\be\label{10.4}
\left(\frac{\dd \T}{\dd\eta}\right)^2=\alpha \T^4+\frac23 \T^3-k\T^2+\frac\lm3, \ee
where
\be
\lm=\frac{\Lambda}{\Lambda_c},\quad \alpha=C_r\Lambda_c,
\ee
are dimensionless parameters called the reduced cosmological constant and the reduced radiation parameter. It is interesting to note \cite{Coq} that the Hubble
parameter now assumes the form
\be
H=\frac{\dd}{\dd t}\ln a=-\sqrt{\Lambda_c}\,\frac{\dd \T}{\dd\eta}.
\ee
When $\alpha=0$, the integration of (\ref{10.4}) may be carried out in terms of elliptic functions \cite{Coq}. Besides, it is clear that, when $\lm=0$ and $\alpha\neq0,k\neq0$, the question whether (\ref{10.4}) can be integrated lies out of the reach of the Chebyshev theorem. Nevertheless, we show here that
the latter case enjoys an exact integration as well as it is given in the cosmic time $t$ and the scale factor $a$.

\subsection{Integration of Equation}

We will only consider the nontrivial case, $\alpha\neq0$. Thus we obtain from (\ref{10.4}) the integral $I=\pm\eta$ where, formally,
\bea
I&=&\int \T^{-1}\left(\alpha \T^2+\frac23 \T-k\right)^{-\frac12}\,\dd \T\nn\\
&=&\int \T^{-1}\left(\alpha \left[\T+\frac1{3\alpha}\right]^2-\left[\frac1{9\alpha}+k
\right]\right)^{-\frac12}\,\dd \T,\label{10.7}
\eea
which spells out another trivial case, $9\alpha k=-1$. So we will only consider the nontrivial case, $9\alpha k\neq -1$, below. Positivity of the quantity under the
square root in (\ref{10.7})
 requires $\alpha>0$ or
\be\label{10.8}
 \frac1{9\alpha}+k<0.
\ee
This observation splits our discussion into a few separate cases.

(i) If $\frac1{9\alpha}+k>0$, then necessarily we have $\alpha>0$ in view of (\ref{10.7}). Hence $\frac19+\alpha k>0$. Now set
\be
\T_{1,2}=\frac{-\frac13\pm\sqrt{\frac19+\alpha k}}\alpha.
\ee
Introduce the new variable $v$:
\be
((\T-\T_1)(\T-\T_2))^{\frac12}=v (\T-\T_1).
\ee
Then $\T=(\T_1 v^2-\T_2)/(v^2-1)$ and 
$
((\T-\T_1)(\T-\T_2))^{\frac12}=-{(\T_2-\T_1)v}/{(v^2-1)}
$.
Then (\ref{10.7}) becomes
\bea
\sqrt{\alpha} I&=&\int \T^{-1}\left((\T-\T_1)(\T-\T_2)\right)^{-\frac12}\,\dd \T\nn\\
&=&-2\int\frac{\dd v}{\T_1v^2-\T_2}.
\eea
If $k=-1$, we have $\T_{2}<\T_1<0$ and 
\be
\T_1 \T_2=\frac1\alpha,\quad \frac{\T_2}{\T_1}=\frac1\alpha\left(\frac13+\sqrt{\frac19-\alpha}\right)^2.
\ee
Hence we obtain
\be\label{x10.13}
I=\ln\left|\frac{3\sqrt{\alpha}\,v-\left(1+\sqrt{1-9\alpha}\right)}{3\sqrt{\alpha}\,v+\left(1+\sqrt{1-9\alpha}\right)}\right|+C,
\ee
where  and in sequel $C$ denotes an integration constant.
If $k=1$, then $\T_2<0<\T_1$ and
\be
\frac{\T_2}{\T_1}=-\frac1\alpha\left(\frac13+\sqrt{\frac19+\alpha}\right)^2.
\ee
So we obtain
\be\label{x10.15}
I=-2\arctan\frac{3\sqrt{\alpha}\,v}{1+\sqrt{1+9\alpha}}+C.
\ee

(ii) If $\frac1{9\alpha}+k<0$, then both subcases, $\alpha>0$ and $\alpha<0$, are allowed. 
Hence, for $\alpha>0$, we have $k=-1$. So, taking $\T+\frac1{3\alpha}=u$, $u=\sqrt{-\frac1\alpha\left(-1+\frac1{9\alpha}\right)}\,\tan\theta$, and $\tan\frac\theta2 =v$, 
consecutively, we obtain
\bea\label{10.13}
I&=&\frac1{\sqrt{\alpha}}\int \T^{-1}\left(\left[\T+\frac1{3\alpha}\right]^2-\frac1\alpha\left[-1+\frac1{9\alpha}\right]\right)^{-\frac12}\,\dd \T\nn\\
&=& 6\sqrt{\alpha}\int\frac{\dd v}{v^2+6\alpha\sqrt{-\frac1\alpha\left(-1+\frac1{9\alpha}\right)}\, v -1}\nn\\
&=&\ln\left|\frac{v-v_1}{v-v_2}\right|+C,
\eea
where  
\be\label{10.18}
v_{1,2}=3\sqrt{\alpha}\left(-\sqrt{1-\frac1{9\alpha}}\pm1\right),
\ee
are the roots of the denominator of the integrand of the second integral in (\ref{10.13}), which further renders (\ref{10.13}) into a more explicit form
\be\label{x10.18}
I=\ln\left|\frac{v+\sqrt{9\alpha-1}-3\sqrt{\alpha}}{v+\sqrt{9\alpha-1}+3\sqrt{\alpha}}\right|+C.
\ee
 Similarly, for $\alpha<0$, both $k=-1$ and $k=1$ are allowed.
So, taking 
$T+\frac1{3\alpha}=u$, $u=\sqrt{\frac1\alpha\left(k+\frac1{9\alpha}\right)}\,\sin\theta$, and $\tan\frac\theta2 =v$, 
consecutively, we obtain
\be\label{10.15}
I={6\sqrt{-\alpha}}\int\frac{\dd v}{v^2-6\alpha\sqrt{\frac1\alpha\left(k+\frac1{9\alpha}\right)}\,v+1}.
\ee
Note that the discriminant of the denominator of the integrand of (\ref{10.15}) is
\be 
\Delta =36\alpha k.
\ee
In the nontrivial situation, $k\neq0$, the condition $\alpha<0$ leads to $\Delta>0$ when $k=-1$ and $\Delta<0$ when $k=1$.
Thus, we arrive at
\be\label{10.21}
I=\left\{\begin{array}{lll} 
&-\ln\left|\frac{v+\sqrt{1-9\alpha}-3\sqrt{-\alpha}}{v+\sqrt{1-9\alpha}+3\sqrt{-\alpha}}\right|+C,\quad &k=-1,\\
&2\arctan\frac{v+\sqrt{9\alpha+1}}{3\sqrt{-\alpha}}+C,\quad & k=1.\end{array}\right.
\ee

It is interesting to notice the similarities between (\ref{x10.13}), (\ref{x10.18}), and the first formula in (\ref{10.21}), and between (\ref{x10.15}) and
the second formula in (\ref{10.21}).

\section{Friedmann Type Equations in Other Contexts}\label{Other}
\setcounter{equation}{0}

It is worth mentioning that the Friedmann type equations are 
also encountered in numerous other physical situations out of the remit  of 
relativistic cosmology.  
Below
we list a  few
examples. By this means one may hope to extend the 
range of the applications of the Chebyshev theorem but
more generally  to gain insight into the
underlying geometry behind the Friedmann equation
and to spot useful ways of representing the solutions.
One particular feature that emerges is that 
many of the solutions may conveniently be represented as roulettes.
This will be the subject of a future paper.
For the present we content ourselves with pointing out 
some correspondences.

We begin by noting that if $k=1$ we have encountered  the
Friedmann equation in three different but equivalent forms
\ben
\frac{1}{a^2} \left(\frac{\dd a }{\dd t} \right) ^2 + \frac{1}{a^2} = \rho(a),  
\een 
\ben
 \left(\frac{\dd a }{\dd \eta } \right) ^2 + a^2  = a^4 \rho(a),  
\een 
\ben
 \left(\frac{\dd\tau  }{\dd \eta } \right) ^2 + \tau ^2  = \rho\left( \frac{1}{\tau}\right),
\een 
where $\dd \eta = a^{-1} \dd t $  and  $\tau=\frac{1}{a}$ .

Note in this section we will  adopt units in which 
$\frac{16\pi G_n}{n(n-1)}=1$.

\subsection{Refraction in a Horizontally Stratified Medium} 

Snell's law for the trajectory of a light ray $y=y(x)$ 
in a horizontally stratified medium reads  
\be\label{7.1}
\left(\frac{y'}y\right)^2+\frac1{y^2}=\frac{n^2}{K^2 y^2},
\ee
where $K>0$ is a constant and $n=n(y)$ is the refractive index.
Equation (\ref{7.1}) is clearly of the $k=1$ Friedmann equation form with
$y(x)$ corresponding to the scale factor $ a(t)$. 
A direct integration gives us
\be \label{7.2}
I=\int\left(\frac{n^2(y)}{K^2}-1\right)^{-\frac12}\,\dd y=\pm x.
\ee

As an example, consider
\be
n(y)=A y^\alpha,
\ee
where $A, \alpha$ are constants, $A\neq0$.
This corresponds to a barotropic equation of state with
$3\gamma= 2(1-\alpha)$. 

 Applying  the Chebyshev theorem, we see that all 
the integrable cases for $\alpha$ are:
\be
\alpha=\frac1N,\quad N=\pm1,\pm2,\dots.
\ee
In particular, dark energy, radiation and dust
correspond to $\alpha=1,-1,-\half$, respectively. Explicitly we have   
\be
I=\left\{\begin{array}{lll}-\frac{A^2}{K^2}\arctan\sqrt{\frac{A^2}{K^2 y}-1}-y\sqrt{\frac{A^2}{K^2 y}-1}+C,&\quad \alpha=-\frac12,\\
-\sqrt{\frac{A^2}{K^2}-y^2}+C,&\quad \alpha=-1,\\
\frac KA\ln\left|\frac AK y+\sqrt{\left(\frac{Ay}K\right)^2-1}\right|+C,&\quad \alpha=1,\end{array}\right.
\ee
where $C$ is an integration constant.

Note that if $\alpha>0 $ the refractive index increases  with height,
the rays are bent upwards, and this  gives rise to a 
typical mirage effect. For example   
in the case of $\alpha=1$, i.e. dark energy, the rays are
 catenary curves. By contrast if $\alpha <0 $, 
the refractive index decreases with height and  the rays are bent downwards.
If $\alpha = -1$, i.e. for radiation,  the rays   are semi-circles and for  
$\alpha = - \half$, i.e. for dust, the rays are cycloids. 
In fact all three curves are roulettes.  The case of dust, $\alpha=-\half$
is equivalent to the brachistochone problem.

Another interesting case  is when the index $n$ in (\ref{7.1})
 assumes the form
\be\label{7.6}
n(y)=\frac {Ay}{1+y},
\ee
where $A$ is a constant satisfying
\be\label{7.7}
\sigma=\frac AK>1. 
\ee
The refractive index increases upwards  and 
the rays are bent upwards.  
Hence the integral $I$ in (\ref{7.2}) becomes
\bea\label{7.8}
I&=&\int\frac{1+y}{\sqrt{(\sigma^2-1)y^2-2y-1}}\,\dd y\nn\\
&=&\frac1{\sqrt{\sigma^2-1}}\int(1+y)\left(\left[y-\frac1{\sigma-1}\right]\left[y+\frac1{\sigma+1}\right]\right)^{-\frac12}\,\dd y,\nn\\
&&\quad\quad  y>\frac1{\sigma-1}\quad \mbox{ or } \quad y<-\frac1{\sigma+1},
\eea
which in view of the Chebyshev theorem may be 
integrated similarly as done in Section 6 and is omitted. 
Here, we simply write down the result:
\be
I=\frac1{\sigma^2-1}\sqrt{(\sigma^2-1)y^2-2y-1}+\frac{\sigma^2}{(\sigma^2-1)^{\frac32}}\ln\left|y-\frac1{\sigma^2-1}+\sqrt{y^2-\frac{2y+1}{\sigma^2-1}}\right|+C,
\ee
where $C$ is an integration constant.

A difficult case is when the index $n$ is given by 
\be
n(y)=\frac{C_0 y^2}{(1+y^2)^2},\quad C_0>0.
\ee
This problem lies beyond the reach of the Chebyshev theorem.

\subsection{Soap Films and Glaciated Valleys}

Soap films are modeled by 
surfaces     
which extremize surface area for fixed volume enclosed.
This is equivalent to their  being of constant mean curvature.
Thus if the soap film is a  surface of revolution
obtained by revolving  the curve $y=y(x)$
about the $x$-axis  we have
\ben
\delta \int  \Bigl( y\sqrt{1 + (y^\prime)^2 } -\lambda y^2\bigr ) \,\dd x=0, 
\een
where $\lambda$  is a Lagrange multiplier enforcing the constant volume
constraint. 
The Euler--Lagrange equation is 
\ben
- \frac{y^{\prime \prime}}{(1+ {y^\prime}^2 ) ^ \frac{3}{2} } +  \frac{1}{
 (1+ {y^\prime}^2 ) ^ \frac{1}{2}} =2 \lambda
\een
which is the statement that the surface has
constant mean curvature since the principal  curvatures
are
\ben
\frac{1}{R_1}= - \frac{y^{\prime \prime}}{(1+ {y^\prime}^2 ) ^ \frac{3}{2} } \,,\qquad \frac{1}{R_2}=\frac{1}{y
 (1+ {y^\prime}^2 ) ^ \frac{1}{2}}.
\een
Noether's theorem gives a first integral 
\ben
- \frac{y}{\sqrt{1+ {y^\prime} ^2 }} + \lambda y^2 = {\rm constant}=c,
\label{constant}
\een
 which may be written in the Friedmann form 
\ben
\left (\frac{y^\prime}{y} \right)^2 + \frac{1}{y^2}= \frac{1}{(c-\lambda y^2) ^2 }\label{Delaunay}.
\een 
  One has the correspondence 
\ben
a \leftrightarrow y\,,\qquad  t  \leftrightarrow x,
    \,, \qquad  \rho \leftrightarrow  
 \frac{1}{(c-\lambda y^2) ^2 }.
\een
Note that in $n$ spatial dimensions, when the $SO(2)$ invariance
is replaced by the $SO(n-1)$ invariance,   $\lambda y^2$ must be replaced 
by $y^{n-1}$ (see \cite{Hsiang}). 
It was discovered by Delaunay  \cite{Delaunay,Eells} that 
we get  in this way     the locus of the focus
of a conic section rolling without slipping on a line. There are three cases, 
the  parabolic catenary, the elliptic catenary, and the hyperbolic catenary.
The surfaces of revolution are called the catenoid, the unduloid and the nodoid 

If $n=3$ and $\lambda=0$ we have a minimal surface of revolution which is well
known to be a catenary. This corresponds to dark energy.
If $n=3$ and $\lambda \ne 0$ the solution may be given  in terms of elliptic 
functions.  If $n>3$ we require in general hypo-elliptic functions \cite{Hsiang}.

The case $c=0$ is rather pathological. 
Superficially, As a soap film it corresponds 
to a cylinder  with $y^\prime =0$. This is a perfectly valid  solution
as a soap film. However its interpretation
as a solution of the Friedmann equation  it would correspond to
a static  closed universe supported by radiation with 
$p= \frac{1}{3} \rho >0$. (This is a well treated case in literature \cite{SK}. See also \cite{CGLY} for a systematic treatment.)
However  $\dot a=0$ does not satisfy the Raychaudhuri equation     
\ben
\frac{\ddot a}{a} = - \frac{4 \pi G}{3} ( \rho + 3P) 
\een
Alternatively if one sets $H=0$ in equations (2,5) and (2,6) 
one obtains a contradiction unless $\frac{1}{3} \rho +P =0$. 

In fact it is easy to show that  the solution
of the $k=1$  Friedmann equation for radiation  in terms of conformal time is
given by  
\ben
a(\eta)=  \frac{1}{\lambda} \cos \eta \,, \qquad t=  \frac{1}{\lambda}
\sin \eta
\een  
which corresponds to a semi-circle in $(x,y)$ space and hence a sphere
when revolved around the $x$ axis.

The  example above was based on 
a variational principle and 
may be further generalized. The Euler--Lagrange equations
of the  variational problem 
\ben
\delta \int  L(y,y^\prime)\,\dd x =0, 
\een  
where
\ben
L= f(y) \sqrt{1+(y^\prime)^2} -g(y) 
\een
admit the first integral
\ben
-\frac{f}{\sqrt{1+(y^\prime)^2} }  + g =c, 
\een
whence 
\ben
\bigl( \frac{y^\prime}{y} \bigr )^2  + 
\frac{1}{y^2} =\frac{f^2}{y^2(c-g)^2}.  
\een

An interesting example is provided by a 
theory of the shape of glaciated valleys \cite{Morgan}
for which $f=y$ and $g=\lambda y$, where the constant $\lambda$
is a Lagrange multiplier. Thus we have
\be 
\left(\frac{y'}{y}\right)^2+\frac1{y^2}=\frac1{(\lm y-c)^2}.
\ee
That is,
\be\label{xy}
(y')^2=\frac{y^2}{(\lm y-c)^2}-1.
\ee
First set
$
u=\lm y-c,
$
assuming $\lm\neq0$ and $c\neq0$, otherwise it is trivial.
Then we have
\be
(u')^2=\frac{(1-\lm^2) u^2 +2c u+c^2}{u^2}.
\ee
Thus we may assume $\lm\neq\pm1$ otherwise it is trivial. With $\lm\neq\pm1$, we use the transformation 
\be
w=(\lm^2-1)\frac yc-\lm
\ee
for comparison with the result in \cite{Morgan} to recast the equation (\ref{xy}) into
\be
(w')^2=\frac{(\lm^2-1)^3}{c^2}\frac{(1-w^2)}{(\lm w+1)^2},
\ee
which can be integrated in view of the Chebyshev theorem. 
Carrying out the integration, we have
\be
\pm\frac{(\lm^2-1)^{\frac32}}{|c|} x=-\lm \sqrt{1-w^2}+\arcsin w +C,\quad w^2<1\quad \mbox{if }\lm^2>1,
\ee
as given in \cite{Morgan}, and
\be
\pm\frac{(1-\lm^2)^{\frac32}}{|c|} x=\lm\sqrt{w^2-1}+\ln\left|w+\sqrt{w^2-1}\right|+C,\quad w^2>1\mbox{ if }\lm^2<1,
\ee
where $C$ is an integration constant.

The  case $c=0$ deserves special treatment.
As a Friedmann equation it  corresponds to a barotropic equation of state with
$\rho>0$ but $\rho+3 P=0$. It thus permits a solution for which $\dot a =0$.
   In general the solution is linear in conformal time. This is consistent
with the equation of state just violates the strong energy condition.

\subsection{Catenary of Equal Strength}

This curve satisfies (\cite{Routh} p. 305 see also \cite{Lamb} p. 192) 
\ben
\frac{y^{\prime \prime}}
{\bigl( 1+ (y^\prime)^2  \bigr ) }= \frac{1}{c}, 
\een 
with first integral
\ben
\arctan y^\prime = \frac{x}{c}+ A, \label{tan}
\een
whence 
\ben
y= B -c \ln \left|\cos \left(\frac{x}{c} +A\right)\right|.\label{why}
\een

Inverting  (\ref{tan})  for $y^\prime$  to get $(y')^2$ as a function
of $\cos^2 (\frac{x}{c} +A )$ and using  (\ref{why})   
lead to
\ben
\frac{1}{y^2}+ \left(\frac{ y^\prime}{y} \right)^2   =
\frac{1}{y^2} \exp\left(\frac{2(y-B)}{c}\right), \label{740}
\een
which is of the Friedmann form under the correspondence
 $(x,y)  \leftrightarrow (t,a)$. Viewed as a closed Friedmann model 
it is time symmetric and exhibits
a ``Big Rip''\cite {Caldwell:1999ew}, i.e a blow up of the scale factor,  at finite time in the future
and the time reverse of a big rip at finite time in the past. However if one evaluates the pressure $P$ the resulting equation of 
state does not seem to be very simple nor to have a simple
physical interpretation. 


 The equation (\ref{740}) is obviously beyond the reach of the Chebyshev theorem.

Identifying the ``matter density" $\rho$ as 
\be\label{741}
\rho= \frac1{y^2}\e^{\frac{2y}c}
\ee
 and applying the ``energy conservation law"
\be\label{xcon}
\rho'+3(\rho+P)\frac{y'}y=0,
\ee
we deduce the equation of state relating the ``pressure" to ``matter density" as follows
\be
P=-\frac13\rho-\frac2{3c}\rho y
=-\frac13\rho-\frac2{3c}\rho f(\rho),
\ee
where $f$ is the inverse function defined by (\ref{741}) if it exists. This equation is complicated. As a by-product, we have
$\rho+P=\frac23\rho(1-\frac yc)=\frac23\rho(1-\frac{f(\rho)}c)$ whose sign is undetermined. However $\rho+3P= -\frac{2\rho y}{c} $ and therefore  
if $y>0$ and $c<0$ we have $\rho +P \ge 0$ and 
$\rho+3P \ge 0$ and both the dominant and the strong energy conditions hold.
On the other hand, if $c>0$ then the strong energy condition 
is violated.

\subsection{The Elastica of Bernoulli and the Capillary Curve} 

Both of these curves satisfy (see \cite{Lamb} p.275) 
\ben
\pm \frac{y^{\prime \prime}} {\bigl (1+ (y^\prime)^2 \bigr ) ^{\frac{3}{2}}}
= \frac{y}{b^2}, \label{sign}
\een
with the first integral
\ben
\mp \frac{1}{\sqrt{1+(y^\prime)^2} } - \frac{y^2}{2b^2} = c. \label{integral}
\een
Turned upside down they give the shape of the ``Hydrostatic Arch''.

Rearranging (\ref{integral})  we have
\ben
\frac{1}{y^2}+ \left(\frac{y^\prime}{y} \right)^2 = \frac{1}{y^2} 
\frac{1}{\bigl( c+\frac{y^2}{2b^2}   \bigr )^2   },\label{744}
\een
which is of the Friedmann form under the correspondence
$(x,y)  \leftrightarrow (t,a)$. 
In general, the solution is given by an elliptic integral 
but if $c=0,\pm1$, then there are some  elementary  solutions (see below).

The  special case $c=0$  has  a linear  equation of state 
corresponding to stiff matter with  positive energy density and 
$ P=\rho$ \footnote{The same
curves arise in the case of the Mylar balloon \cite{Paulsen}.
If one interchanges $x$ and $y$ one obtains (\ref{integral}) 
and the Mylar balloon corresponds to setting our integration constant
$c$ to zero. More generally, any equation of the form
$  \frac{y^\prime} {\sqrt{1+ (y^\prime )^2  } }  = f(x)$
  may be converted to the Friedmann equation form in this way.} 
The Friedmann equation in terms of conformal time 
becomes 
\ben
\left( \frac{\dd a}{\dd \eta} \right) ^2 + 4 a^4 = 16 b^4, 
\een
 whence
\ben
a= a(0)  \sqrt{\cos (2 \eta)} \,.
\een 
Cosmic time is then given by an elliptic integral. 

If $c=-1$ and we take the upper sign in (\ref{integral}) we can find 
a complete solution for $y$ and $x$ in elementary terms.
For a direct evaluation using the integrals  see below.
In order to better understand the global 
behavior of the solutions,  we first recall  the 
traditional approach (see e.g.  \cite{Lamb}) 
and we will take contact with the discussion based on direct integration
below. 
We define $\psi$ to be $y^\prime = \tan \psi$, so $\psi$ 
is the angle the curve makes with the $x$-axis.  
the first integral (\ref{integral})  then becomes 
\ben
c \mp \cos \psi = \frac{y^2}{2b^2}. \label{trig}
\een
Setting $c=-1$ and taking the upper sign in (\ref{trig}) 
we have 
\ben
y= 2b \sqrt{\sin\left(\frac{\psi}{2}\right) }\,, \label{y}
\een
where we have chosen the positive sign for the square root
and we shall ultimately  be interested in $\psi$ ranging from $0$ to $2 \pi$. 
Now
\ben
\frac{\dd x}{\dd \psi} = \frac{\dd x}{\dd s} \frac{\dd s}{\dd \psi},
\een
where $\dd s=\sqrt{1+(y^\prime)^2 }\,\dd x $ along the curve,
and using the fact that the curvature of  a curve is given by
\ben
\frac{y^{\prime \prime}}
{\bigl( 1+ (y^\prime)^2  \bigr) ^{\frac32}} = \frac{\dd \psi}{\dd s }, 
\een
and (\ref{why}) we find that 
\ben
\frac{\dd x}{\dd \psi} = 
\pm \half b \frac{\cos \psi}{\sin \frac{\psi}{2} } = \pm
\half b \left( \frac{1} {\sin\frac{\psi}{2} } - 2 \sin\frac{\psi}{2}  
\right).\label{deriv}
\een
Taking the upper sign in (\ref{deriv}) we see that $x$ increases with 
increasing $\psi$
from $\psi=0$ to $\psi = \frac{\pi}{2} $, it then decreases 
with increasing $\psi$ until $ \psi= \frac{3 \pi}{2}$ and thereafter increases
with increasing $\psi$ until $\psi=2 \pi$.    
Integrating (\ref{deriv}) we find cosmic time to be given by  
\ben
t= x= b \ln\left(\tan \frac{\psi}{4}\right ) + 2 b \cos \frac{\psi}{2}. 
\een
We see that  as $\psi$ runs from $0$ to $2 \pi$, $x$ runs from $-\infty$
to $+\infty$,  and the curve, which is symmetric about
the $y$-axis    has a node on the $y$-axis and 
a loop running from the node  and back  thereafter continue to infinity.
Between $\psi=\frac{\phi}{2}$ and  $\psi=\frac{3 \pi}{2}$ the function
$y(\psi)$  is double valued.  On the branch from $\psi=0$ to $\psi = \pi$ the function
$y(\psi)$  is monotonic increasing from zero  and on  the branch from    
$\psi=\pi $ to $\psi =2 \pi$  it is monotonic decreasing to zero. 
The maximum value of $y$ is $2b$. 
At the node $y= 2 b \sin\frac{\psi_n}{2}$ where  $\psi_n$ satisfies
\ben
\ln\left(\tan \frac{\psi_n }{4}\right ) + 2  \cos \frac{\psi_n}{2}=0.
\een 
   
Considered  as a closed Friedmann model conformal time is given by
\ben
\dd \eta = \frac{\dd x}{y}= \frac{\dd t}{a} =  \frac{1}{4} \left( 
\frac{1}{\sin ^2\frac{\psi}{2} }
  -2  \right) \dd \psi.
\een 
That is, with a choice of integration constant, 
\ben
\eta = - \half \cot \left(\frac{\psi}{2}  +\psi -\pi   \right).
\een

If we take the branch that runs from $t=-\infty$ to $ t=0$ we obtain
a cosmological model expanding from zero size in the infinite past and 
reaching a finite size  in finite time at $t=0$ at the node.
The question then  is what happens next? 
The continuation past the singular point is not obvious.
If joined to the lower branch, then Hubble's constant
$H= \frac{\dot a}{a}$  would jump from
a positive to an equal negative value at $t=0$.  
However the Einstein  equations would be violated
and the jump in Hubble's constant  would
correspond to a spacelike hypersurface on which some
matter instantaneously appears from nowhere.   

A better alternative  would be to follow  the loop.
The question is then what happens when one  reaches $\psi =\frac{\pi}{2}$? 
Hubble's constant  is given by
\ben
H= \frac{1}{y} \frac{\dd y}{\dd \psi} \frac { \dd \psi}{\dd x} 
= \frac{1}{\cos \frac{\psi}{2}} 
\, \frac{\sin ^2 \frac{\psi}{2}}
{(1-2 \sin^2 \frac{\psi}{2} ) },  
\een
which diverges as $\psi \rightarrow \frac{\pi}{2} $.
This seems to be an example of a sudden singularity \cite{Barrow:2004xh}.

From (\ref{744}) we can read off the ``matter density"  to be
\be
\rho=\frac1{y^2\left(c+\frac{y^2}{2b^2}\right)^2}.
\ee
Hence with $H=\frac{y'}y$ we have
\be
\rho=-2H\left(1+\frac{y^2}{\left(b^2 c+\frac12 y^2\right)}\right)\rho. 
\ee
Comparing this result with the conservation law (\ref{xcon}), we obtain
\be
\frac13\rho+P=\frac{2\rho y^2}{3\left(b^2 c+\frac12 y^2\right)},
\ee
which is positive for any $c\geq0$. This is the strong energy condition, which implies the dominant energy condition
$\rho+P>0$. If $c<0$, the sign of $\rho+P$ is generally undetermined.


One might also consider the top portion of the loop.
This gives a time-symmetric closed universe  which   starts
with a big bang at finite size $y=\sqrt2 b$,
with an infinite Hubble constant and  
reaches a time of maximum  expansion at a finite time thereafter 
after which it   recollapses to a big crunch.  

In both of the two cases just discussed,  the singularity
is clear from (\ref{744}) since if $c=-1$ the energy  
density diverges as $y\rightarrow \sqrt2 b$ as 
$\frac{1}{(y-\sqrt2 b)^2} $.  This is reminiscent of
the behavior of an equation of state with $P=-\frac{1}{3} \rho$ . 

Similar possibilities are possible for solutions
for which $c\ne -1$.  In particular if $c$ is negative similar
singularities can arise. If $c>0$ the behavior  is more like the $c=0$
case. 

In general, setting 
\be
u=\frac{y^2}{2b^2}+c
\ee
in (\ref{744}) leads formally to the integration
\be
\pm\frac{\sqrt{2}}{|b|}\, x=\int\frac{u\,\dd u}{\sqrt{(u-c)(1-u^2)}}\equiv I_c,
\ee
which lies beyond the reach of the Chebyshev theorem unless $c=0,\pm1$. So it is worthwhile to boil down the range of $c$ in which the problem is relevant. For this
purpose, we rearrange (\ref{744}) to get
\be
\left(c+\frac{y^2}{2b^2}\right)^2(y')^2=\left([1-c]-\frac{y^2}{2b^2}\right)\left(1+c+\frac{y^2}{2b^2}\right).
\ee
So we see that the value $c>1$ is prohibited and $c=1$ gives rise to
 the trivial solution, $y\equiv0$. For the remaining nontrivial integrable case, $c=-1$, we have
\bea
I_{-1}&=&\sqrt{2}\,\mbox{arctanh}\sqrt{\frac{1-u}2}-2\sqrt{1-u}+C\nn\\
&=&\sqrt{2} \,\mbox{arctanh}\sqrt{1-\frac{y^2}{4b^2}}-2\sqrt{2}\sqrt{1-\frac{y^2}{4b^2}}+C,
\eea
where $C$ is an integration constant.
To make contact with our previous discussion we set 
\ben
\sqrt{1-\frac{y^2}{4b^2}} = \cos \frac{\psi}{2} 
\een
and use the identity 
\ben
\mbox{arctanh}\cos \frac{\psi}{2}= \ln \cot \frac{\psi}{2}.  
\een
\subsection{Central Orbits} 
Consider a particle moving in a plane with polar coordinates
$r,\theta$. 
If $v$ is the speed and $V(r)$ the potential energy  per unit mass
then by  energy conservation,   
\ben
\half \bigl( \dot r^2 + r^2 \dot \theta ^2 \bigr )  + V(r) = \cE,
\een
where $\cE$ is the energy per unit mass. 
Angular momentum conservation gives
\ben
r^2 \dot \theta = h, 
\een
where $h$ is the angular momentum per unit mass.
Thus
\ben
\frac{1}{r^4} \left(\frac{\dd a }{\dd \theta} \right)^2  +  \frac{1}{r^2}
= - \frac{2}{ h^2}  \bigl ( V(r)-\cE \bigr ).  
\een
Now setting $u=\frac{1}{r}$ 
we obtain
\ben
\left(\frac{\dd u}{\dd  \theta} \right)^2 + u^2 =  
- \frac{2}{h^2}  \left( V\left(\frac{1}{u} \right)-\cE \right).  
\een

We see that this equation is of the same form as the reduced temperature version
of the Friedmann equation (\ref{10.4}) with $k=1$, 
under the correspondence $(u, \theta)
\leftrightarrow ( \tau , \eta)  $ and bearing in mind that while
$\theta$ is a periodic coordinate period $2 \pi$, conformal time $\eta$
is not periodic.

Central orbit problems  arise from considering geodesics
in spherically symmetric metrics. In this connection
it is perhaps of interest that the equation (\ref{10.4}) 
for the reduced temperature is of the same form as that obtained   
for null geodesics in the Reissner--Nordstrom metric (see
equation (51) in  \cite{Gibbons:2011rh})
except the sign of the cosmological term is negative
for real values of the electric charge. 

\subsection{Spherical  Symmetric  Lenses} 

A closely related example to the previous one 
is provided by applying Fermat's principle to 
a lens whose refractive index $n(r)$ depends on the
radial coordinate  $r$.  
The motion may be assumed to be  in a plane through the origin and we have
\ben
n^2(r) \Bigl( {\dot r}  ^2 + r^2 {\dot \theta}^2 \Bigr )  = 1, 
\een 
and 
\ben
n^2 r^2 \dot \theta = h =  n r \sin i,
\een
where $h$ is a constant and $i$ is the inclination between
the ray and the radial direction.

Thus 
\ben
\left(\frac{\dd r}{r^2  \dd \theta}\right) ^2 + \frac{1}{r^2} = \frac{n^2}{h^2}   
\een
and so if $u=\frac{1}{r}$ we have 
\ben
\left(\frac{\dd u}{ \dd \theta}\right) ^2 +u^2 =  \frac{n^2(\frac{1}{u}) }{h^2}. 
\een 

Note that while the lens problem is defined in the plane
and hence $\theta$ is periodic of period $2 \pi$, no such restriction
is placed on $t$. Thus the cosmological problem is defined 
on the universal covering space of the punctured plane. 

Some examples are as follows:

\begin{itemize}
\item Maxwell's fish-eye lens is given by
\ben
n(r) = \frac{n_0}{1+r^2}, 
\een
and thus
\ben
\rho= \frac{n^2_0}{h^2 (1+a^2)^2 }.
\een
Comparing with (\ref{Delaunay}) and Delauny's construction
of soap films,    
it would appear that this case corresponds to the 
locus of the focus of an ellipse rolled without
slipping along a line (i.e $\lambda c <0$). Now the optical metric of
Maxwell's fish eye lens is that of a round 3-sphere
pulled back to ${\Bbb R}^3$ by stereographic projection
and hence all rays are circles. If these circles
enclose the origin and lie in a plane through the origin
they presumably map to the elliptic catenaries 
on the covering space.
\item
The Eaton lens has 
\ben
n= \sqrt{\frac{2a}{r}-1}, \qquad   0<r\le a;\qquad n=1,\qquad  r>a. 
\een
Thus for $r<a$,  the  solutions of which are portions of ellipses
\ben
\frac{l}{r}= 1+e \cos (\phi-\phi_0) 
\een
with focus at the center $r=0$, semi-latus rectum
$l$,  and eccentricity $e$, where
\ben
l=\frac{h^2}{a},\qquad e= \sqrt{1-\frac{b^2}{a^2}}.
\een  
The ends of the semi-major axes of an elllipse are situated at a radial
 distance
\ben
\frac{l}{1-e^2} 
\een
from the focus at $r=0$. In our case one finds that this distance
is $a$-independent of the impact parameter $h$. 
Because the refractive index is continuous across the boundary, rays 
entering the lens initially continue in the same direction
are thus  initially at one end of the semi-minor axis and 
initially moving perpendicularly  to the semi-minor axis of their elliptical
trajectory  within the lens. It follows that they exit 
from the other end of the semi-minor axis  in the opposite direction from
that  which  enter the lens.   Thus the Eaton lens behaves like
a  retroreflector: {\sl Any} ray entering the lens exits in the
opposite direction.   

\item

As pointed out in \cite{Hannay} the Eaton lens
is dual in the sense of Arnold--Bohlin to the Luneburg lens
for  which   
\ben
n= \sqrt{2-\frac{r^2}{a^2} },\qquad r\le a;\qquad n=1,\qquad r>a.
\een
\end{itemize}

\section{Conclusion}

In this paper we have discussed the explicit integration
of Friedmann's equations
for equations of state more complicated than that of a single  
component fluid  of 
linear barotropic form. In particular we have made use
of a theorem of Chebyshev which gives sufficient conditions
for explicit integrability. However we also found 
examples, while which are outside the scope of Chebyshev's
result, nevertheless allow explicit integration.
As well as the traditional  form of the Friedmann equation    is 
in terms of cosmic time, we have treated the case when the independent
variable is conformal time. Less traditionally we have also
considered the inverse of the scale factor (the so-called
reduced temperature) as a function of conformal time.  
We have seen through a thorough study of the Chaplygin fluid model that integrable cases may often provide sharp insight into nonintegrable cases
as well. In particular,
based on some integrable examples, we have derived a universal formula for the asymptotic exponential growth rate of the scale factor in cosmic time, regardless
whether the Friedmann equation falls in its integrable regime, and the formula reveals the coupled roles played by the conventional linear matter and nonconventional
nonlinear matter, with regard to the presence of dark matter.
Finally we have shown that all three forms of the Friedmann
equation occur in  areas of physics not directly connected with cosmology.
This suggests that methods of dealing with the equation
developed in those areas may transfer to the cosmological setting.
In particular in a forthcoming paper we will  illustrate 
this point by expressing its  solutions as roulettes.

\medskip

The research of Chen was supported in part by
Henan Basic Science and Frontier Technology Program
Funds under Grant No. 142300410110.  Yang was partially supported by National Natural Science Foundation of China under Grant No. 11471100. 

\medskip

\small{

}
\end{document}